\documentclass[12pt]{article}
\usepackage{graphicx}
\usepackage{float} %Include figure filesusepackage{graphicx} %Include figure files
\usepackage{amsmath}
\usepackage{slashed}
\usepackage{amsfonts}   
\usepackage{amssymb}

\begin{document}
\date{}
%%%%%%%%%%%%%%%%%%%%
\title{{\bf{\Large Models of phase stability in Jackiw-Teitelboim gravity}}}
%%%%%%%%%%%%%%%%%%%%
\author{{\bf {\normalsize Arindam Lala}$
$\thanks{E-mail:  arindam.physics1@gmail.com, arindam.lala@pucv.cl}}\\
 {\normalsize  Instituto de F\'{i}sica, Pontificia Universidad Cat\'{o}lica de Valpara\'{i}so,}\\
  {\normalsize Casilla 4059, Valparaiso, CHILE,}
\\[0.3cm]
 {\bf {\normalsize Dibakar Roychowdhury}$
$\thanks{E-mail:  dibakarphys@gmail.com, dibakarfph@iitr.ac.in}}\\
 {\normalsize  Department of Physics, Indian Institute of Technology Roorkee,}\\
  {\normalsize Roorkee 247667 Uttarakhand, India}
\\[0.3cm]
}
%\date{}

\maketitle
%%%%%%%%%%%%%%%%%%%%%%%%%%%%%%%%%%%%%%%%%%%%%%%%%%%%%%%%%%%%%%%%%%%%%%%%%%%%
\begin{abstract}
We construct solutions within Jackiw-Teitelboim (JT) gravity in the presence of
nontrivial couplings between the dilaton and the Abelian $1$-form where we analyse the 
asymptotic structure as well as the phase stability corresponding to charged black hole
solutions in $1+1$ D. We consider the Almheiri-Polchinski (AP) model as a specific example 
within $1+1$ D JT gravity which plays a pivotal role in the study of Sachdev-Ye-Kitaev (SYK)/anti-de Sitter (AdS) duality. The 
corresponding vacuum solutions exhibit a rather different asymptotic structure than their 
uncharged counterpart. We find interpolating vacuum solutions with AdS$_2$ in the IR and 
Lifshitz$_2$ in the UV with dynamical exponent $z_{\text{dyn}}=3/2$. Interestingly, the 
presence of charge also modifies the black hole geometry from asymptotically AdS to 
asymptotically Lifshitz with same value of the dynamical exponent. We consider specific 
examples, where we compute the corresponding free-energy and explore the thermodynamic
phase stability associated with charged black hole solutions in $1+1$ D. Our analysis 
reveals the existence of a universal thermodynamic feature that is expected to reveal
its immediate consequences on the dual SYK physics at finite density and strong coupling.
\end{abstract}
%%%%%%%%%%%%%%%%%%%%%%%%%%%%%%%%%%%%
\section{Introduction and Motivations}
For the last couple of decades, there had been considerable efforts towards a profound understanding of the
underlying non-perturbative dynamics in large $N$ gauge theories using the celebrated AdS/CFT framework
\cite{Maldacena:1997,Witten:1998,Aharony:2000}. Nonetheless, till date there exists only a few examples
where this duality can actually be tested with precise accuracy. In other words, one can exactly solve the
spectrum on both sides of the duality albeit they are strongly interacting. In the recent years, an example of this 
kind has emerged where the spectrum of the $0+1$ dimensional strongly interacting Sachdev-Ye-Kitaev (SYK) 
model can be solved exactly using large $N$ techniques whose dual counterpart has been conjectured to be the 
Jackiw-Teitelboim (JT) model in $1+1$ D \cite{Sachdev:1992fk}-\cite{Moitra:2019}. Apart from being 
exactly solvable, the SYK model exhibits maximal chaos together with an emergent conformal symmetry at low 
energies which therefore provides a reliable platform to test the holographic correspondence. 

For the last couple of years, there has been a systematic effort towards unveiling the dual gravitational counterpart 
of the SYK model. A hint came from the JT dilaton gravity in $1+1$ D \cite{Teitelboim:1983ux}-\cite{Grumiller:2014oha}
based on which disparate dual gravity models have been proposed \cite{Almheiri:2014cka}-\cite{Das:2017wae} along with 
several interesting extensions \cite{Yoshida1:2017}-\cite{Roychowdhury:2018}.

The original SYK/AdS duality deals with Majorana fermions for which neutral dual gravity models are enough to 
consider. This has been the line of analyses for most of the models so far. However, very recently charged SYK 
models have been constructed in \cite{Davison:2016ngz,Gaikwad:2018dfc} whose dual gravitational counterpart
has been proposed to be given by the 2D \emph{effective} gravity action of the following form\footnote{See 
Appendix \ref{dim:redn} for details.}\cite{Davison:2016ngz},

%%%%
\begin{equation}\label{action:gen}
S_{2D}\sim \int d^{2}x \sqrt{-g}\left(\Phi^{2} \mathcal{R}+V(\Phi^{2})
-\frac{Z(\Phi^{2})}{4}F^{\mu\nu}F_{\mu\nu}\right).
\end{equation}
%%%%

The last term in the above action (\ref{action:gen}) represents the non-trivial coupling between the dialton and the
Abelian $1$-form. This interaction term can be viewed as an effective coupling which can be obtained as a result of 
dimensional reduction from the $3+1$ D version of the theory \cite{Davison:2016ngz}. Eq. (\ref{action:gen}) without 
the gauge field is precisely the form of the action considered in \cite{Almheiri:2014cka} with the potential $V(\Phi^{2})$ 
linear in dilaton. In the present analysis, we choose to work with two specific forms of the dilaton coupling, namely 
$Z(\Phi^{2})\sim\left(\Phi^{2}\right)^{2}$ and $Z(\Phi^{2})\sim e^{-\Phi^{2}}$ together with the choice of the dilaton 
potential $V(\Phi^{2})$ as given in \cite{Almheiri:2014cka}.

Based on the classical gravity computations, we construct charged 2D black hole solutions in the two aforementioned 
models. Our analysis reveals that the presence of charge substantially modifies the asymptotic symmetries of the 
space-time, namely converting it to a two dimensional asymptotic Lifshitz geometry which otherwise would have been an 
AdS$_2$ geometry. We further compute the free-energy and explore the thermodynamic phase stability of the 
obtained solutions. For both the models, we observe a \textit{universal} thermodynamic feature of phase stability at sufficiently
low temperatures and finite density.

The organisation of the paper is as follows: In Section \ref{formulation}, we propose our first model with $Z(\Phi^{2})
\sim\left(\Phi^{2}\right)^{2}$. Considering linear potential for the dilaton potential we explore the bounds on the potential 
which makes the space-time asymptotically AdS. In Sections \ref{flat:sol} and \ref{grav:ads2} we comment on the 
vacuum structures as well as the black hole solutions both for an asymptotically flat and asymptotically AdS
space-time, respectively. In Section \ref{sol:gen} we construct perturbative solutions (in charge, $Q$) to our model and 
analyse the underlying geometry associated to both the vacuum and as well as the charged black hole solutions. This 
is supplemented with the study of the phase stability of derived solutions using the standard background subtraction 
method \cite{Witten:1998bs}. In Section \ref{exp:grav} we repeat our analysis for the exponential dilaton coupling. Finally, 
we conclude in Section \ref{conclusions} where we mention about the possible implications of our findings on the corresponding
SYK counterpart.

%%%%%%%%%%
\section{Example I: Quadratic coupling}\label{formulation}
We start with the Einstein-Maxwell-dilaton action of the following form,
%%%%
\begin{equation}
\label{act:charged}
S=-\int d^{2}x \sqrt{-g}\left(\mathcal{R}\Phi^{2}-U(\Phi^{2})-\frac{1}{4}
\left(\Phi^{2}\right)^{2}F^{2}\right)-\int dt \sqrt{-\gamma}\;\Phi^{2}\mathcal{K}
\end{equation}
%%%%
where $F=dA$ is the Maxwell 2-form field, $\Phi$ is the dilaton and $U(\Phi^{2})$ is the dilaton potential. Notice that, we
have added the the Gibbons-Hawking-York boundary term \cite{York:1972,Gibbons:1977} in the above action, where 
$\gamma$ is the determinant of the induced metric on the boundary and $\mathcal{K}$ is the trace of the extrinsic 
curvature \cite{Natsuume:2015}. In the subsequent analysis we set the AdS length scale $L=1$ and $16\pi G=1$.
%\footnote{In our analysis, we shall choose a $z=\text{Const.}$ hypersurface. The outward pointing unit normal 
%$n^{z}$ is then defined as, $g_{zz}n^{z}n^{z}=1$. The trace of the extrinsic curvature can subsequently be written as, 
%$\displaystyle\mathcal{K}=n^{z} \frac{\partial_{z}\sqrt{-\gamma}}{\sqrt{-\gamma}}$.} 

The equations of motion can be written as,
%%%%
\begin{subequations}
\setlength{\jot}{8pt}
\begin{eqnarray}
0&=&\left(\nabla_{\mu}\nabla_{\nu}-g_{\mu\nu}\Box\right)\Phi^{2}+\frac{1}{2}\Phi^{4}\left(F_{\mu\rho}
F_{\nu}^{\;\;\rho}-\frac{1}{4}F^{2}g_{\mu\nu}\right)-\frac{1}{2}g_{\mu\nu}U(\Phi^{2}) \label{a:metric}\\
0&=&\mathcal{R}-\frac{F^{2}}{2}\Phi^{2}-\frac{\partial U(\Phi^{2})}{\partial\Phi^{2}} \label{b:scalar}\\
0&=&\partial_{\mu}\left(\sqrt{-g}\Phi^{4}F^{\mu\nu}\right)  \label{c:gauge}.
\end{eqnarray}
\end{subequations}
%%%%
Let us now consider the conformal gauge
%%%%
\begin{equation}
\label{gauge:conf}
ds^{2}=-e^{2\omega(x^{+},x^{-})}dx^{+}dx^{-}
\end{equation}
%%%%
where $\displaystyle x^{\pm}=t\pm z$. In this light-cone gauge the equations of motion can be written as,
%%%%
\begin{subequations}\label{eom:lc}
\setlength{\jot}{8pt}
\begin{eqnarray}
4\partial_{+}\partial_{-}\Phi^{2}-e^{2\omega}U(\Phi^{2})-2\Phi^{4}
e^{-2\omega}F_{+-}^{2}&=&0 \label{a:eom}\\
\partial_{\pm}\partial_{\pm}\Phi^{2}-2\left(\partial_{\pm}\omega\right)
\left(\partial_{\pm}\Phi^{2}\right)&=&0 \label{b:eom}\\
4\partial_{+}\partial_{-}\omega -\frac{e^{2\omega}}{2}\frac{\partial 
U(\Phi^{2})}{\partial \Phi^{2}}+2e^{-2\omega}\Phi^{2}
F_{+-}^{2}&=&0 \label{c:eom}\\
\partial_{\pm}\chi&=&0 \label{d:eom}
\end{eqnarray}
\end{subequations}
%%%%
where we have defined $\chi=\sqrt{-g}\;\Phi^{4}F^{+-}$. Notice that our analysis differs from that of \cite{Almheiri:2014cka} 
in the sense that there is no fully decoupled equation of motion for $\omega(z)$. As a result we obtain a different solution 
for the conformal factor $e^{2\omega(z)}$.

In the next step we would like to consider the static solutions. In order to do so, we revert back to the $(t,z)$ coordinates
and use the following ansatz for the gauge field,
%%%%
\begin{align}\label{anz:gf}
%\begin{split}
A_{+}&=\frac{1}{2}\left(A_{t}+A_{z}\right) \qquad \qquad  A_{-}=\frac{1}{2}\left(A_{t}-A_{z}\right)
\qquad \qquad A_{\mu}=\left(A_{t},0\right).
%\end{split}
\end{align}
%%%%

In these coordinates (\ref{a:eom})-(\ref{d:eom}) can be expressed as,
%%%%
\begin{subequations}
\setlength{\jot}{8pt}
\begin{eqnarray}
 (\Phi^{2})''+e^{2\omega}U(\Phi^{2})+\frac{1}{2}\Phi^{4}e^{-2\omega}A_{t}'^{2}&=&0\label{a:steom}\\
(\Phi^{2})''-2\omega'(\Phi^{2})'&=&0 \label{b:steom}\\
 2\omega''+e^{2\omega}\frac{\partial U(\Phi^{2})}{\partial \Phi^{2}}-e^{-2\omega}\Phi^{2}A_{t}'^{2}&=&0
\label{c:steom}\\
\Phi^{2}A_{t}''-2A_{t}'\left(\Phi^{2}\omega'-(\Phi^{2})'\right)&=&0 \label{d:steom}.
\end{eqnarray}
\end{subequations}
%%%%

We can rewrite (\ref{a:steom}) in the following form
%%%%
\begin{equation}
\label{const:phia}
U(\Phi^{2})=-e^{-2\omega}\left(\Phi^{2}\right)''-\frac{1}{2}\Phi^{4}e^{-4\omega}A_{t}'^{2}.
\end{equation}
%%%%

Let us now consider a constant dilaton profile, $\Phi^{2}(z)=\Phi^{2}_{0}$. In this case substituting (\ref{const:phia}) in
(\ref{c:steom}) we obtain
%%%%
\begin{equation}
\label{const:wa}
2\omega''=-2e^{2\omega}\frac{\partial U(\Phi^{2})}{\partial \Phi^{2}}\Bigg|_{\Phi^{2}=\Phi^{2}_{0}}.
\end{equation}
%%%%

The Ricci scalar ($\mathcal{R}$) on the other hand, can be written as
%%%%
\begin{equation}
\label{ricci}
\mathcal{R}=2\frac{\partial U(\Phi^{2})}{\partial \Phi^{2}}\Bigg|_{\Phi^{2}=\Phi^{2}_{0}}.
\end{equation}
%%%%
where we have used (\ref{const:wa}).

Now, in order for the space-time to have AdS$_2$ asymptotics we must have the following criteria:
%%%%
\begin{equation}
\label{cond:curv}
\frac{\partial U(\Phi^{2})}{\partial \Phi^{2}}\Bigg|_{\Phi^{2}=\Phi^{2}_{0}}<0.
\end{equation}
%%%%

\section{Gravity in asymptotically flat $1+1$ D}\label{flat:sol}
\subsection{Minkowski vacuum}\label{min:vs}
In this section, we would like to consider the vacuum solution corresponding to $1+1$ D dilaton gravity. In order to do
so we consider the following metric
%%%%
\begin{equation}\label{metric:vsmin}
ds^{2}=-dt^{2}+dz^{2}.
\end{equation}
%%%%

From the gauge equation of motion (\ref{c:gauge}) we can write
%%%%
\begin{equation}\label{gf:vsmin}
F_{tz}=\frac{Q}{\Phi^{4}}.
\end{equation} 
%%%%

It is easy to check that the scalar equation of motion (\ref{b:scalar}) and one of the trace equations of motion corresponding 
to the metric, (\ref{a:metric}), lead to the following form of the dilaton potential,
%%%%
\begin{equation}\label{pot:vsmin}
U(\Phi^{2})=-\frac{1}{2}\frac{Q^{2}}{\Phi^{4}}.
\end{equation}
%%%%

On the other hand, the remaining trace equation of motion can be written expressed as,
%%%%
\begin{equation}\label{ttmet:vsmin}
\left(\Phi^{2}\right)''+\frac{1}{4}\frac{Q^{2}}{\Phi^{4}}+\frac{1}{2}U(\Phi^{2})=0.
\end{equation}
%%%%

Substituting (\ref{pot:vsmin}) into (\ref{ttmet:vsmin}) we obtain
%%%%
\begin{equation}\label{dileqn:vsmin}
\left(\Phi^{2}\right)''=0
\end{equation}
%%%%
whose solution may be written as,
%%%%
\begin{equation}\label{dilaton:vsmin}
\Phi^{2}=bz+a
\end{equation}
%%%%
where $a$ and $b$ are arbitrary integration constants.

%%%%%%%%%%%%%%%%%%%%%%
%%%%%%%%%%%%%%%%%%%%%%

\subsection{Charged black hole solutions}\label{chg:bh}
We choose to work with metric ansatz of the following form,
%%%%
\begin{equation}
\label{metric:ansatz}
ds^{2}=-f(z)dt^{2}+f^{-1}(z)dz^{2}.
\end{equation}
%%%%

The only component of the Maxwell field strength tensor can thus be written as
%%%%
\begin{equation}
\label{max:comp}
F_{tz}=\frac{Q}{\Phi^{4}(z)}
\end{equation}
%%%%
where $Q$ is an integration constant which we can identify as the charge of the black hole, and we have used (\ref{c:gauge})
in order to derive (\ref{max:comp}).

With the metric (\ref{metric:ansatz}) the remaining equations of motion (\ref{a:metric}) and (\ref{b:scalar}) can be wriiten as,
%%%%
\begin{align}
f'(z)\left(\Phi^{2}\right)'+\frac{Q^{2}}{2\left(\Phi^{2}\right)^{2}}+U(\Phi^{2})&=0 \label{metzz:cor}\\
2f(z)\left(\Phi^{2}\right)''+f'(z)\left(\Phi^{2}\right)'+\frac{Q^{2}}{2\left(\Phi^{2}\right)^{2}}+U(\Phi^{2})
&=0  \label{mettt:cor}\\
f''(z)-\frac{Q^{2}}{\Phi^{6}(z)}+\frac{\partial U(\Phi^{2})}{\partial\Phi^{2}}&=0.  \label{scal:cor}
\end{align}
%%%%

From (\ref{metzz:cor}) and (\ref{mettt:cor}) it is easy to check that 
%%%%
\begin{equation}\label{phieom}
\left(\Phi^{2}\right)''=0
\end{equation}
%%%%
and as a result the dilaton becomes constant at the boundary $z=0$.

As a next step, we determine the metric coefficient $f(z)$ with particular choices of the dilaton potential $U(\Phi^{2})$
where we finally set, $\Phi^{2}=\varphi_{0}=\text{const}$.

%%%%
\begin{flushleft}
\textbf{Case I:} \underline{The Almheiri-Polchinski model} \cite{Almheiri:2014cka} ($U(\Phi^{2})=C-A\Phi^{2}$) and 
\underline{the Callan-Giddings-Harvey-Strominger (CGHS) model} \cite{Callan:1992rs} ($U(\Phi^{2})=-A\Phi^{2}$)
\end{flushleft}
%%%%

From (\ref{scal:cor}) we obtain,
%%%%
\begin{equation}\label{metf:eom}
f''(z)=A+\frac{Q^{2}}{\varphi_{0}^{3}}
\end{equation}
%%%%
whose solution may be written as,
%%%%
\begin{align}\label{fsol:apc}
\begin{split}
f(z)=\left(1-\frac{z}{z_{H}}\right)\left[1-\frac{1}{2}\left(A+\frac{Q^{2}}{\varphi_{0}^{3}}
\right)zz_{H}\right].
%&=c_{1}+zc_{2}+\frac{z^{2}}{2}\left(A+\frac{Q^{2}}{\varphi_{0}^{3}}\right)
\end{split}
\end{align}
%%%%

Using (\ref{fsol:apc}), the Hawking temperature of the black hole can be obtained as,
%%%%
\begin{align}\label{temp:AP}
\begin{split}
T_{H}&=\frac{1}{4\pi}\partial_{z}\sqrt{-\frac{g_{tt}}{g_{zz}}}\Bigg|_{z=z_{H}}  \\
&=\frac{1}{4\pi z_{H}}\left[\frac{1}{2}\left(A+\frac{Q^{2}}{\varphi_{0}^{3}}\right)
z_{H}^{2}-1\right].
\end{split}
\end{align}
%%%%

The extremal limit, in which the Hawking temperature vanishes, is characterized by the extremal value
of the charge given by,
%%%%
\begin{equation}\label{ext:AP}
Q_{e}^{2}=\left(\frac{2}{z_{H}^{2}}-A\right)\varphi_{0}^{3}.
\end{equation}
%%%%

On the other hand, by using the Wald formalism, the entropy of the black hole is given by \cite{Myers:1994},
%%%%
\begin{equation}\label{ent:AP}
S_{W}=4\pi \Phi^{2}.
\end{equation}
%%%%

The thermodynamic stability of the black hole is determined by computing the corresponding heat capacity ($C$),
%%%%
\begin{equation}\label{sh:AP}
C=T_{H}\frac{\partial S_{W}}{\partial T_{H}}=\frac{8\pi \varphi_{0}^{4}}{3Q^{2}z_{H}^{2}}
\left[1-\frac{1}{2}\left(A+\frac{Q^{2}}{\varphi_{0}^{3}}\right)z_{H}^{2}\right]
\end{equation}
%%%%
where we have used (\ref{temp:AP}) and (\ref{ent:AP}).

For non-extremal black holes one must have $T_{H}> 0$ which leads to the following condition,
%%%%
\begin{equation}\label{temp:ext}
\frac{1}{2}\left(A+\frac{Q^{2}}{\varphi_{0}^{3}}\right)z_{H}^{2}>1.
\end{equation}
%%%%
Substituting (\ref{temp:ext}) in (\ref{sh:AP}) we note that the specific heat is always negative. This suggests that 
the black holes (\ref{fsol:apc}) in an asymptotically flat space-time are indeed unstable and decay through Hawking
radiation. 

%%%%
\begin{flushleft}
\textbf{Case II:} \underline{Magnetic branes} \cite{DHoker:2009}, $U(\Phi^{2})=\frac{B^{2}}{\Phi^{2}}-A\Phi^{2}$
\end{flushleft}
%%%%

The equation of motion corresponding to the metric $f(z)$ can be expressed as,
%%%%
\begin{equation}\label{fmet:mb}
f''(z)=A+\frac{B}{\varphi_{0}^{2}}+\frac{Q^{2}}{\varphi_{0}^{3}}
\end{equation}
%%%%
where we have used (\ref{scal:cor}). The general solution to the above equation (\ref{fmet:mb}) is given by,
%%%%
\begin{align}\label{fsol:mb}
\begin{split}
f(z)=\left(1-\frac{z}{z_{H}}\right)\left[1-\frac{1}{2}\left(A+\frac{B}{\varphi_{0}^{2}}+\frac{Q^{2}}{\varphi_{0}^{3}}
\right)zz_{H}\right].
%&=c_{3}+zc_{4}+\frac{z^{2}}{2}\left(A+\frac{B}{\varphi_{0}^{2}}+\frac{Q^{2}}{\varphi_{0}^{3}}\right) \\
\end{split}
\end{align}
%%%%

Proceeding in the same line of analysis as in the previous Case I, the thermodynamic quantities may be found 
as follows:
%%%%
\begin{itemize}

\item The Hawking temperature: \begin{equation}\label{temp:MB}
													T_{H}=	\frac{1}{4\pi z_{H}}\left[\frac{1}{2}\left(A+\frac{B}{\varphi_{0}^{2}}
													+\frac{Q^{2}}{\varphi_{0}^{3}}\right)z_{H}^{2}-1\right].
													\end{equation}
													
\item Extremal value of charge:  \begin{equation}\label{extq:MB}
													Q_{e}^{2}=\left[\frac{2}{z_{H}^{2}}-\left(A+\frac{B}{\varphi_{0}^{2}}\right)
													\right]\varphi_{0}^{3}.
													\end{equation}		
													
\item The specific heat: 
%%%%
\begin{align}\label{sh:MB}
\begin{split}
C&=T_{H}\frac{\partial S_{W}}{\partial T_{H}}\\
&=\frac{4\pi\varphi_{0}^{3}}{z_{H}^{2}}\left[1-\frac{1}{2}\left(A+\frac{B}{\varphi_{0}^{2}}
+\frac{Q^{2}}{\varphi_{0}^{3}}\right)z_{H}^{2}\right]\left(B+\frac{3Q^{2}}{2\varphi_{0}}\right)^{-1}.
\end{split}
\end{align}																			
%%%%
Like in the previous example, it is easy to check that the condition for extremality implies the thermodynamic
instability in black holes. 

\end{itemize}
%%%%

%%%%%%%%%%%%%%%%%%%%%%%%%%%%
%%%%%%%%%%%%%%%%%%%%%%%%%%%%

\section{Vacuum solutions with AdS$_2$ asymptotics}\label{grav:ads2}
Unlike the previous example, here we discuss the possibilities on vacuum solutions with AdS$_2$ asymptotics. We
show that solutions with AdS$_2$ asymptotics are indeed possible both for the constant as well as the running dilaton
profiles.
\subsection{Solution with constant dilaton}
We first construct solutions with constant dilaton $\Phi^{2}=\Phi_{0}^{2}$.
\subsubsection{Case I: $e^{2\omega}=\frac{1}{z^{2}}$}\label{case1}

In this case from (\ref{a:steom}) we may write
%%%%
\begin{equation}
\label{gauge}
A_{t}'^{2}=-\frac{2}{z^{4}\Phi_{0}^{4}}U(\Phi_{0}^{2})
\end{equation}
%%%%
whereas (\ref{b:steom}) is satisfied trivially. Notice that, in the above equation $U(\Phi_{0}^{2})$ must be negative. Using the 
relation $\displaystyle \omega'=-\frac{1}{z}$ we can write (\ref{c:steom}) as,
%%%%
\begin{equation}
\label{potder}
\frac{2}{z^{2}}+\frac{1}{z^{2}}\frac{\partial U(\Phi^{2})}{\partial \Phi^{2}}\Bigg|_{\Phi_{0}^{2}}-
z^{2}\Phi_{0}^{2}A_{t}'^{2}=0.
\end{equation}
%%%%

Finally, using (\ref{gauge}) we obtain the following relation:
%%%%
\begin{equation}
\label{potder:rel}
\frac{\partial U(\Phi^{2})}{\partial \Phi^{2}}\Bigg|_{\Phi_{0}^{2}}=-2\left(1-\frac{|U(\Phi_{0}^{2})|}{
\Phi_{0}^{2}}\right)
\end{equation}
%%%%
which leads to the following bound to the potential
%%%%
\begin{equation}
\label{potbound}
\frac{|U(\Phi_{0}^{2})|}{\Phi_{0}^{2}}<1.
\end{equation}
%%%%

\subsubsection{Case II: $e^{2\omega}=\frac{1}{\sinh^{2}z}$}

In this case the equation for the gauge field can be expressed as,
%%%%
\begin{equation}
\label{gauge2}
A_{t}'^{2}=-\frac{2}{\Phi_{0}^{4}\sinh^{4}z}U(\Phi_{0}^{2})
\end{equation}
%%%%
from which we again conclude that $U(\Phi_{0}^{2})<0$. Substituting $e^{2\omega}$ in (\ref{c:steom}) it is trivial to check that 
one arrives at the same constraint conditions (\ref{potder:rel}) and (\ref{potbound}). 

\subsubsection{Case III: $e^{2\omega}=\frac{1}{\sin^{2}z}$}

In this case the equation for the gauge field can be written as,
%%%%
\begin{equation}
\label{gauge:der}
A_{t}'^{2}=-\frac{2}{\Phi_{0}^{4}\sin^{4}z}U(\Phi_{0}^{2})
\end{equation}
%%%%
from which it is evident that $U(\Phi_{0}^{2})<0$. Substituting $e^{2\omega}$ in (\ref{c:steom}) we once again arrive at the 
same constraint conditions (\ref{potder:rel}) and (\ref{potbound}).

Thus we observe that the above constraint conditions (\ref{potder:rel}), (\ref{potbound}) are universal for all the three cases
considered, i.e. they are same irrespective of the choice of the coordinate systems.

%%%%%%%%%%%%%%%%%%%%%%
%%%%%%%%%%%%%%%%%%%%%%
\subsection{Solutions with running dilaton}
%\subsubsection{Vacuum AdS$_2$ solutions}\label{ads2:vs}
Let us consider the metric of the following form,
%%%%
\begin{equation}\label{metric:vsa1}
ds^{2}=\frac{1}{z^{2}}\left(-dt^{2}+dz^{2}\right).
\end{equation}
%%%%

With this choice of metric the gauge equation of motion (\ref{c:gauge}) leads to,
%%%%
\begin{equation}\label{gf:vsa1}
F_{tz}=\frac{Q}{\Phi^{4}z^{2}}.
\end{equation}
%%%%

The metric equation of motion can be expressed as,
%%%%
\begin{align}
z\left(\Phi^{2}\right)'-\frac{1}{4}\frac{Q^{2}}{\Phi^{4}}-\frac{1}{2}U(\Phi^{2})&=0 
\label{met1:vsa1}\\
z^{2}\left(\Phi^{2}\right)''+z\left(\Phi^{2}\right)'+\frac{1}{4}\frac{Q^{2}}{\Phi^{4}}
+\frac{1}{2}U(\Phi^{2})&=0.\label{met2:vsa1}
\end{align}
%%%%

Substituting (\ref{met1:vsa1}) into (\ref{met2:vsa1}) we obtain,
%%%%
\begin{equation}\label{deq:vsa1}
z^{2}\left(\Phi^{2}\right)''+2z\left(\Phi^{2}\right)'=0
\end{equation}
%%%%
whose solution may be formally expressed as,
%%%%
\begin{equation}\label{dsol:vsads1}
\Phi^{2}=-\frac{b_{1}}{z}+a_{1}.
\end{equation}
%%%%
It is interesting to note that the dilaton diverges near the boundary, $z\approx 0$.

In a similar way, considering the conformal factor as
$e^{2\omega}=\sinh^{-2}z,\;\sin^{-2}z$, the solutions to the dilaton
can be found as,
%%%%
\begin{align}
\Phi^{2}&=-b_{2}\coth z+a_{2}\label{dil:a} \\
\Phi^{2}&=-b_{3}\cot z +a_{3},\label{dil:b}
\end{align}
%%%%
respectively. Interestingly, both these solutions diverge near the boundary, $z\approx 0$. 

Notice that, the divergence of the dilaton profile (\ref{dsol:vsads1})-(\ref{dil:b}) near the boundary ($z\approx 0$) is a generic
feature of the AdS space-time, see for example \cite{Apruzzi:2014qva} and references therein. However, as far as the present 
analysis is concerned, one of the solutions (\ref{dsol:vsads1}) has an interesting consequence from the perspective of the 
SYK/AdS duality. One of the interesting facets of this duality is that it allows us to relate the SYK degrees of freedom to the
underlying dynamics of the AdS$_2$ counterpart \cite{Maldacena:2016hyu},\cite{Das:2017pif}-\cite{Das:2017wae}. We can
consistently set $a_{1}=0$ in (\ref{dsol:vsads1}) by demanding that the dilaton $\Phi^{2}$ vanishes as we probe deep IR ($z
\rightarrow\infty$). This observation is in fact consistent with the strongly interacting ($J\sim 1/\Phi^{2}\gg1$) nature of the dual 
SYK model in which we are mostly interested \cite{Das:2017pif}-\cite{Das:2017wae}).

%The SYK degrees of freedom correspond to infrared (IR) dynamics on the AdS$_2$ counterpart. From 
%(\ref{dsol:vsads1}) it is clear that the dilaton ($\Phi^{2}\sim 1/J$ \cite{Das:2017pif}-\cite{Das:2017wae}) vanishes (modulo a 
%trivial constant term) as we probe deep IR ($z\rightarrow\infty$) dynamics on the gravity side. This observation is in fact 
%consistent with the strongly interacting ($J\gg1$) nature of the dual SYK model in which we are mostly interested.

%%%%%%%%%%%%%%%%%%%%%%
%%%%%%%%%%%%%%%%%%%%%%
\section{General solutions: A perturbative approach}\label{sol:gen}
In this section, we adopt perturbation techniques in order to find the general solutions to the equations of motion 
(\ref{a:metric})-(\ref{c:gauge}) and determine the metric of the space-time. In order to perform our analysis, we 
consider a running dilaton where the dilaton is a function of $z$ only: $\Phi^{2}=\Phi^{2}(z)$. We also consider 
the dilaton potential as\footnote{See Appendix \ref{sols:neg} regarding black hole solution with $A<0$.} 
\cite{Almheiri:2014cka},
%%%%
\begin{equation}\label{pot:mod1}
U(\Phi^{2})=C-A\Phi^{2}, \qquad \quad C,A> 0.
\end{equation}
%%%%

In order to obtain solutions to the metric as well as the dilaton equations of motion we expand the above entities
as a perturbation in the $U(1)$ charge $Q$ namely,
%%%%
\begin{subequations}
\setlength{\jot}{8pt}
\begin{eqnarray}
\Phi^{2}(z)&=&\Phi^{2}_{(0)}(z)+Q^{2}\Phi^{2}_{(1)}(z)+\cdots \label{solper:phi}  \\
\omega(z)&=&\omega_{(0)}(z)+Q^{2}\omega_{(1)}(z)+\cdots \label{solper:om}
\end{eqnarray}
\end{subequations} 
%%%%

The Maxwell field strength tensor is given by the solution of (\ref{d:steom}) which may be written as,
%%%%
\begin{equation}
\label{maxwell:ns}
F_{zt}\equiv A_{t}'=-\frac{Qe^{2\omega}}{\Phi^{4}(z)}.
\end{equation}
%%%%

On the other hand, the equation of motion for the dilaton (\ref{b:steom}) can be recast in the following form
%%%%
\begin{equation}
\label{dilaton:rd}
(\Phi^{2})'=\tilde{C}\;e^{2\omega}
\end{equation}
%%%%
whose solution may be written as,
%%%%
\begin{equation}\label{dila:rd}
\Phi^{2}(z)=\tilde{C}\int e^{2\omega} dz+\bar{C}
\end{equation}
%%%%
where $\tilde{C}$, $\bar{C}$ are arbitrary integration constants. 
%In the subsequent analysis
%we shall fix the constants $\bar{C}$ using the equations of motion.

In the following we note down the equations of motion upto leading order in the perturbative expansion. Substituting 
(\ref{solper:om}) into (\ref{c:steom}) we find, 
%%%%
\begin{align}
\mathcal{O}(Q^{0}):\qquad 0&=\omega_{(0)}''-e^{2\omega_{(0)}} \label{Q0:om}\\
\mathcal{O}(Q^{2}):\qquad 0&=2\omega_{(1)}''-4e^{2\omega_{(0)}}\omega_{(1)}-
e^{2\omega_{(0)}}\left(\Phi^{2}_{(0)}\right)^{-3}. \label{Q1:om}
\end{align}
%%%%

On the other hand, substituting (\ref{solper:phi}) into (\ref{a:steom}) we find,
%%%%
\begin{align}
\mathcal{O}(Q^{0}):\qquad 0&= \left(\Phi_{(0)}^{2}\right)''+2e^{2\omega_{(0)}}
\left(1-\Phi_{(0)}^{2}\right) \label{Q0:phi}\\
\mathcal{O}(Q^{2}):\qquad 0&=\left(\Phi_{(1)}^{2}\right)''-2e^{2\omega_{(0)}}\Phi_{(1)}^{2}
+4e^{2\omega_{(0)}}\left[\omega_{(1)}\left(1-\Phi_{(0)}^{2}\right)+\frac{1}{8} 
\left(\Phi_{(0)}^{2}\right)^{-2}\right]. \label{Q1:phi}
\end{align}
%%%%

%%%%%%%%%%%%%%%%%%%%%%%%
%%%%%%%%%%%%%%%%%%%%%%%%
\subsection{Interpolating vacuum solutions}\label{vac:sol}
The vacuum solution corresponding to $Q=0$ is characterised by \cite{Almheiri:2014cka}
%%%%
\begin{align}
e^{2\omega_{(0)}^{\text{vac}}(z)}&=\frac{1}{z^{2}}  \label{vsol:w0} \\
\Phi_{(0)}^{2^{\text{vac}}}(z)&=\left(1+\frac{1}{2z}\right)   \label{vsol:p0}
\end{align}
%%%%
which satisfy the zeroth order equations (\ref{Q0:om}) and (\ref{Q0:phi}), respectively for the following values of the 
constants in (\ref{dila:rd}): $\tilde{C}=-1/2$ and $\bar{C}=1$. Thus for the purpose of our present analysis it is 
sufficient to solve the $\mathcal{O}(Q^2)$ equations of motion namely, (\ref{Q1:om}) and (\ref{Q1:phi}).

Using (\ref{Q1:om}) the solution corresponding to $\omega_{(1)}^{\text{vac}}$ can be expressed as,\footnote{We
set the second integration constant, $\mathsf{C}_{3}$, to zero. This is to render the on-shell
action in Section \ref{phtran} finite at the horizon.}
%%%%
\begin{equation}\label{vsol:w1}
\omega_{(1)}^{\text{vac}}(z)=\frac{\mathsf{C}_{4}}{z}+\frac{1-2z-4z^{2}
+2(1+2z)\cdot\log(1+2z)}{8z(1+2z)}.
\end{equation} 
%%%%

Thus the metric for the vacuum can be expressed as,
%%%%
\begin{equation}\label{vsol:met}
ds^{2}=e^{2\omega_{(0)}}\left(1+2Q^{2}\omega_{(1)}^{\text{vac}}\right)
\left(-dt^{2}+dz^{2}\right).
\end{equation}
%%%%

Let us now analyse the IR and UV behaviors of the solution (\ref{vsol:met}).
%%%%
\begin{itemize}

\item In the IR region $z\rightarrow\infty$ the behavior of the metric is given by,
%%%%
\begin{equation}\label{vmet:ir}
e^{2\omega}\simeq \frac{1}{z^{2}}\left(1-\frac{Q^{2}}{2}
\right)+\mathcal{O}\left(z^{-3}\right)
\end{equation}
%%%%
which thereby leads to an \emph{emerging} AdS$_2$ geometry. 

\item The UV ($z=0$) behavior of the metric is found to be of the following form,
%%%%
\begin{equation}\label{vmet:uv}
e^{2\omega}\simeq \frac{Q^{2}}{4z^{3}}\left(1+8\mathsf{C}_{4}\right)
+\frac{1}{z^{2}}-\frac{2Q^{2}}{3}+\mathcal{O}(z)
\end{equation}
%%%%
whose leading contribution comes from the first term on the R.H.S of (\ref{vmet:uv}). This turns out to be a 
Lifshitz$_2$ geometry with dynamical exponent $z_{\text{dyn}}=\frac{3}{2}$.

\end{itemize}
%%%%

Thus we observe that the vacuum solution interpolates between Lifshitz$_2$ in the UV and AdS$_2$ in the deep IR. 
It is interesting to note that in the absence of the charge, $Q=0$, the geometry is AdS$_2$ for both in the IR as well
as the UV. Thus we conclude that the presence of the gauge field modifies the UV asymptotics from AdS$_2$ to 
Lifshitz$_2$ \cite{Taylor:2008}. 

%%%%%%%%%%%%%%%%%%%%%%
%%%%%%%%%%%%%%%%%%%%%%

\subsection{Charged 2D black holes}\label{sol:lif2}

The zeroth-order solutions $\Phi^{2}_{(0)}(z)$ and $\omega_{(0)}(z)$ are respectively given by 
\cite{Almheiri:2014cka},
%%%%
\begin{subequations}
\setlength{\jot}{8pt}
\begin{eqnarray}
\Phi^{2}_{(0)}&=&1+\sqrt{\mu}\coth(2\sqrt{\mu}z) \label{solper:phi0}  \\
e^{2\omega_{(0)}}&=&\frac{4\mu}{\sinh^{2}(2\sqrt{\mu}z)} \label{solper:om0}
\end{eqnarray}
\end{subequations} 
%%%%
Notice that, in order for the solution (\ref{solper:phi0}) to be consistent with the equations of motion (\ref{a:steom}) 
and (\ref{b:steom}) we must choose the constants appearing in (\ref{dila:rd}) as $\tilde{C}=-1/2$ and $\bar{C}
=1$. 

It is trivial to check that, with the choice of the constants $\tilde{C}$ and $\bar{C}$, (\ref{solper:phi0}) and 
(\ref{solper:om0}) are indeed the solutions to the equations of motion (\ref{Q0:om}) and (\ref{Q0:phi}). 
Using (\ref{solper:phi0}) and (\ref{solper:om0}) in (\ref{Q1:om}) the solution for $\omega_{(1)}^{BH}$ may be 
written as,
%%%%
\begin{align}\label{om1:sol}
\begin{split}
&\omega_{(1)}^{BH}(\rho)  \\
&=\mathsf{C}_{1}\frac{\rho}{\sqrt{\mu}}+\mathsf{C}_{2}\left[\frac{\rho}{2\sqrt{\mu}}
\cdot\log\left(\frac{\sqrt{\mu}+\rho}{\sqrt{\mu}-\rho}\right)-1\right]+\frac{1}{8\mu^{3/2}
\left(\mu-1\right)^{2}}\times   \\
&\qquad\Bigg\{-2\sqrt{\mu}\left(1-\mu\frac{2+\rho}{1+\rho}+\frac{\mu^{2}}{1+\rho}\right)+
4\mu^{3/2}\rho\cdot\log(1+\rho)  \\
&\qquad -\rho\cdot\log(\rho-\sqrt{\mu})\left(1-3\mu+2\mu^{3/2}\right)
+\rho\cdot\log(\rho+\sqrt{\mu})\left(1-3\mu-2\mu^{3/2}\right)\Bigg\}
\end{split}
\end{align}
%%%%
where $\mathsf{C}_{1}$, $\mathsf{C}_{2}$ are the integration constants, and we have used the following change
in the spatial coordinate \cite{Almheiri:2014cka,Yoshida1:2017}
%%%%
\begin{equation}\label{coor:new}
z\longrightarrow\frac{1}{2\sqrt{\mu}}\coth^{-1}\left(\frac{\rho}{\sqrt{\mu}}\right).
\end{equation}
%%%%
In the subsequent calculations we set $\mathsf{C}_{2}=0$ in order to obtain a physically meaningful asymptotic
structure of the space-time. 

Finally, using (\ref{solper:om}), (\ref{solper:om0}) and (\ref{om1:sol}) the metric (\ref{gauge:conf}) corresponding to 
the black hole geometry can be expressed as,
%%%%
\begin{align}\label{met:Apos}
\begin{split}
ds^{2}&=e^{2\omega}\left(-dt^{2}+dz^{2}\right)\\
&=\frac{4\mu}{\sinh^{2}2\sqrt{\mu}z}\left(1+2Q^{2}\omega_{(1)}^{BH}(z)\right)
\left(-dt^{2}+dz^{2}\right)\\
&=4\left(\rho^{2}-\mu\right)\left(1+2Q^{2}\omega_{(1)}^{BH}(\rho)\right)\left(-dt^{2}+
\frac{d\rho^{2}}{4\left(\rho^{2}-\mu\right)^{2}}\right).
\end{split}
\end{align}
%%%%

Notice that, the above black hole solution (\ref{coor:new}) has a horizon at $\rho=\sqrt{\mu}$. On the other hand, 
the boundary is located at $\rho=\infty$.  Expanding the metric near the boundary we find,
%%%%
\begin{equation}\label{sol:bdy}
e^{2\omega(\delta)}\simeq \frac{8Q^{2}\mathsf{C}_{1	}}{\sqrt{\mu}\delta^{3}}+\frac{4}{\delta^{2}}
-\frac{8Q^{2}\mathsf{C}_{1}\sqrt{\mu}}{\delta}-4\mu+\mathcal{O}(\delta)
\end{equation}
%%%%
where we have changed the variable, $\displaystyle \rho\rightarrow1/\delta$ and taken the limit $\delta\rightarrow 0$ 
subsequently. The leading term in the expansion (\ref{sol:bdy}) behaves as $\displaystyle\sim\frac{1}{\delta^{3}}$ 
which is a signature of an asymptotically Lifshitz geometry with dynamical critical exponent $z_{\text{dyn}}=\frac{3}{2}$. 
It is trivial to check that in the limit $Q^{2}\rightarrow 0$ the resulting metric is that of an asymptotically AdS$_2$ 
space-time. Our result thus indicates that in the presence of the gauge field the asymptotic behaviour of the space-time 
indeed changes from AdS$_2$ to Lifshitz$_2$ \cite{Taylor:2008}. 

Next, we turn our attention towards computing the dilaton profile for our model. Using (\ref{dila:rd}) we observe that,
%%%%
\begin{align}\label{sol:phi1}
\begin{split}
\Phi_{(1)}^{2}&=-\int e^{2\omega_{0}}(z)\omega_{(1)}(z)dz \\
&=2\int \omega_{(1)}(\rho) d\rho.
\end{split}
\end{align} 
%%%%

Notice that, in writing the second line we have used (\ref{coor:new}). Thus the complete solution upto 
$\mathcal{O}(Q^{2})$ can be expressed as,
%%%%
%%%%
\begin{align}\label{tot:phi}
\begin{split}
\Phi^{2}&\simeq\Phi_{(0)}^{2}+Q^{2}\Phi_{(1)}^{2} \\
&=(1+\rho)+2Q^{2}\int \omega_{(1)}(\rho) d\rho.
\end{split}
\end{align} 
%%%%

We now compute the thermodynamic quantities corresponding to the above black hole geometry (\ref{met:Apos}). 
The corresponding Hawking temperature is given by,
%%%%
\begin{align}
\label{temp:bh}
\begin{split}
T_{H}&=\frac{1}{2\pi}\sqrt{-\frac{1}{4}g^{tt}g^{\rho\rho}\left(\partial_{\rho}
g_{tt}\right)^{2}}\Bigg|_{\rho\rightarrow\sqrt{\mu}}=\frac{\sqrt{\mu}}{\pi}.
\end{split}
\end{align}
%%%%

On the other hand, the Wald entropy associated with the black hole can be found as, 
%%%%
\begin{align}\label{ent:bh}
\begin{split}
S_{W}&=4\pi\left[\left(1+\sqrt{\mu}\right)+2Q^{2}\int\omega_{(1)}(\rho)d\rho
\Big|_{\rho=\sqrt{\mu}}\right]  \\
&=4\pi\left[1+\sqrt{\mu}+Q^{2}\frac{1+4\mathsf{C}_{1}\mu|\mu-1|}{4\sqrt{\mu}|\mu-1|}\right]
\end{split}
\end{align}
%%%%
Notice that (\ref{ent:bh}) is ill defined at $\mu=1$. Therefore one can define black hole solutions for either
of the two branches namely, $\mu<1$ or $\mu>1$ which is thereby consistent with the earlier observations 
\cite{Almheiri:2014cka,Yoshida1:2017}.

%%%%%%%%%%%%%%%%%%%%%%%%
%%%%%%%%%%%%%%%%%%%%%%%%

\subsection{Phase stability}\label{phtran}
In order to check whether there is any phase transition/crossover between the empty AdS$_2$ and the AdS$_2$ black 
hole one needs to compare free-energies between different configurations. We substitute (\ref{b:scalar}) into the action 
(\ref{act:charged}) which finally yields,
%%%%
\begin{equation}\label{act2}
S=-\int d^{2}x \sqrt{-g}\left(\frac{F^{2}}{4}\left(\Phi^{2}\right)^{2}-C\right)
-\int dt \sqrt{-\gamma}\;\mathcal{K}\Phi^{2}.
\end{equation}
%%%%

Next we use the equation of motion for the gauge field, (\ref{c:gauge}) to find the on-shell action as,
%%%%
\begin{align}\label{act:os}
\begin{split}
-S^{(os)}&=-\int d^{2}x \sqrt{-g}\;C+\int d^{2}x \;\partial_{\mu}\left(\sqrt{-g}\;F^{\mu\nu}
\left(\Phi^{2}\right)^{2}A_{\nu}\right)  \\
&=-C\cdot \text{(Vol)} + \int dt \; \left(\sqrt{-g}\;F^{zt}
\left(\Phi^{2}\right)^{2}A_{t}\right)+\int dt \sqrt{-\gamma}\;\mathcal{K}\Phi^{2}
\end{split}
\end{align}
%%%%
where $\displaystyle \text{(Vol)}$ is the volume term associated with the following two geometries:

(1) For empty interpolating vacuum we may write down the volume term as,
%%%%
\begin{align}\label{vol:emp}
\begin{split}
\left(\text{Vol}\right)_{\text{Int}}^{\text{vac}}&=-\int_{0}^{\beta_{0}}d\tau
\int_{0}^{\Lambda}d\rho\left(\frac{1}{\rho^{2}}\right)
e^{2\omega_{\text{Int}}^{\text{vac}}}(\rho)     \\
&=-\beta_{0}\;\mathcal{A}_{\text{vac}}(\rho)\Big|_{0}^{\Lambda}
\end{split}
\end{align}
%%%%
where we have used the change of coordinate $\displaystyle z\rightarrow \frac{1}{\rho}$ and
%%%%
\begin{equation}\label{int:vac}
\mathcal{A}_{\text{vac}}=\rho-\frac{Q^{2}\rho}{2}\left(1-\frac{\rho}{4}\left(1+8
\mathsf{C}_{4}\right)-\frac{\rho}{2}\cdot\log\left(\frac{2+\rho}{\rho}\right)\right).
\end{equation}

(2) For the AdS$_2$ black hole the volume term can be expressed as,
%%%%
\begin{align} \label{vol:bh}
\begin{split}
\left(\text{Vol}\right)^{BH}&=-\int_{0}^{\beta_{1}}
d\tau\int_{\rho\sim\sqrt{\mu}}^{\rho\sim\Lambda}d\rho \frac{1}{2
(\rho^{2}-\mu)} e^{2\omega^{\text{BH}}}(\rho)  \\
&=-\beta_{1}\;\left(\mathcal{A}^{(\Lambda)}(\rho\rightarrow\Lambda)
-\mathcal{A}^{(\mu)}(\rho\rightarrow\sqrt{\mu})\right)
\end{split}
\end{align}
%%%%
where,
%%%%
\begin{align}\label{term:cut}
\begin{split}
&\mathcal{A}^{(\Lambda)}(\rho\rightarrow\Lambda)  \\
&=\frac{1}{2\mu^{3/2}(\mu-1)^{2}}\Big\{2\Lambda\sqrt{\mu}(\mu-1)\Big[2\mu(\mu-1)
+Q^{2}\left(1+\Lambda\mathsf{C}_{1}\sqrt{\mu}(\mu-1)\right)\Big]  \\
&\quad+Q^{2}(\Lambda^{2}-\mu)\Big[4\mu^{3/2}\cdot\log(1+\Lambda)+(-1+3\mu-2\mu^{3/2})
\cdot\log(\Lambda-\sqrt{\mu}) \\
&\quad+(1-3\mu-2\mu^{3/2})\cdot\log(\Lambda+\sqrt{\mu})\Big]
\Big\}
\end{split}
\end{align}
and
\begin{equation}\label{term:hori}
\mathcal{A}^{(\mu)}(\rho\rightarrow\sqrt{\mu})=\frac{4\mu(\mu-1)+Q^{2}\Big[1+4\mu
\mathsf{C}_{1}(\mu-1)\Big]}{2\sqrt{\mu}(\mu-1)}.
\end{equation}
%%%%
Here, we have introduced an UV cut-off $\Lambda$  in order to make the integrals finite, and $\beta_{0}$ and 
$\beta_{1}$ are the periods associated with the interpolating vacuum and the AdS$_2$ black hole, respectively. 
Moreover, while $\beta_{1}$ is fixed by (\ref{temp:bh}), $\beta_{0}$ is arbitrary. Thus we can fix $\beta_{0}$ by 
demanding that at some arbitrary $\rho=\Lambda$, the temperature of both the configurations should be the 
same namely,\footnote{ The choice of normalization (\ref{beta:rat}) for $\beta_{0}$ is \emph{unique} in the sense that
the geometry of the hypersurface placed at a radial cut-off $\rho=\Lambda (\rightarrow \infty)$ should be same both for 
the vacuum as well as the black hole space-time \cite{Witten:1998conf}. This also ensures the consistency in the thermodynamic description in the sense of holography as the limit $\Lambda\rightarrow\infty$ is approached. }
%%%%
\begin{equation}\label{beta:rat}
\frac{\beta_{0}}{\beta_{1}}=\sqrt{\frac{e^{2\omega^{\text{BH}}}(\rho)}
{e^{2\omega_{\text{Int}}^{\text{vac}}}(\rho)}}\Bigg|_{\rho\rightarrow\Lambda}.
\end{equation}
%%%%

In the next step, we would like to compute the difference between the volume terms (\ref{vol:bh}) and (\ref{vol:emp}).
This may be written in the following form,
%%%%
\begin{align}\label{dif:vol}
\begin{split}
\Delta\left(\text{Vol}\right)&=-C\left(\left(\text{Vol}\right)^{BH}
-\left(\text{Vol}\right)_{\text{Int}}^{\text{vac}} \right)   \\
&=C\beta_{1}\Big[\left(\mathcal{A}_{1}-\mathcal{B}_{1}\right)\Lambda^{2}
+\left(\mathcal{A}_{2}-\mathcal{B}_{2}\right)\Lambda+
\left(\mathcal{A}_{3}-\mathcal{B}_{3}\right)-\mathcal{A}^{(\mu)}(\rho
\rightarrow\sqrt{\mu})\Big]
\end{split}
\end{align}
%%%%
where the individual coefficients are given by
%%%%
\begin{align}\label{cofs:A}
\mathcal{A}_{1}=\frac{2Q^{2}\mathsf{C}_{1}}{\sqrt{\mu}}, \qquad\quad \mathcal{A}_{2}=2,  
\qquad\quad \mathcal{A}_{3}=\frac{Q^{2}}{2(\mu-1)}
\end{align}
%%%%

%%%%
\begin{align}\label{cofs:B}
\begin{split}
\mathcal{B}_{1}&=\frac{Q^{2}}{\sqrt{2}}\sqrt{\frac{\mathsf{C}_{1}(1+8\mathsf{C}_{4})}{\sqrt{\mu}}} \\
\mathcal{B}_{2}&=\frac{1}{4\sqrt{2}\mathsf{C}_{1}}\sqrt{\frac{\mathsf{C}_{1}}{\sqrt{\mu}(1+8\mathsf{C}_{4})}}
\left(\sqrt{\mu}\left(1+8\mathsf{C}_{4}\right)+24\mathsf{C}_{1}\right)  \\
\mathcal{B}_{3}&=-Q^{2}\left(1+\mathsf{C}_{1}\sqrt{\mu}\right).
\end{split}
\end{align}
%%%%

It is to be noted that in (\ref{cofs:B}) we have simplified the cumbersome expression for $\mathcal{B}_{3}$ 
by using the following relation,
%%%%
\begin{equation}\label{rel:cont}
\mathsf{C}_{1}=\frac{\sqrt{\mu}}{8}\left(1+8\mathsf{C}_{4}\right)
\end{equation}
%%%%
which is easily derived by setting the coefficient of $\Lambda^{2}$ in (\ref{dif:vol}) equal to zero. In addition, using
(\ref{rel:cont}), it is easy to check that the coefficient of the $\Lambda$ term in (\ref{dif:vol}) vanishes. 

Let us now consider the second term $\Pi$ in (\ref{act:os}). Using (\ref{maxwell:ns}), (\ref{met:Apos}) and 
(\ref{vsol:met}) this may be written as,
%%%%
\begin{align}\label{cor:vol}
\begin{split}
\Pi&=Q^{2}\int d\tau\int_{0}^{\Lambda}e^{2\omega_{(0)}(z)}
\left(\Phi_{(0)}^{2}(z)\right)^{-2}dz \\
&=\begin{cases}-4\beta_{0}Q^{2}\left(\frac{1}{2+\rho}\right)\Big|_{0}^{\Lambda}&\quad
\text{(interpolating vacuum)}\\[20pt]
-2\beta_{1}Q^{2}\left(\frac{1}{1+\rho}\right)\Big|_{\sqrt{\mu}}^{\Lambda}&\quad 
\text{(black hole)}\end{cases}
\end{split}
\end{align}
%%%%
where we have substituted (\ref{solper:phi0}) and (\ref{solper:om0}). Finally, in the limit $\Lambda\rightarrow
\infty$ the difference between the $\text{2$^{nd}$ terms}$ can be written as,
%%%%
\begin{equation}\label{dif:2nd}
\Delta\left(\Pi\right)=\Pi^{\text{BH}}-\Pi^{\text{vac}}=-2Q^{2}\beta_{1}
\left(\frac{1+2\sqrt{\mu}}{1+\sqrt{\mu}}\right).
\end{equation}
%%%%

We now calculate the contribution from the third term $\Sigma$ in (\ref{act:os}). In order to do so, we choose a $\rho=
\text{Const.}$ hypersurface \cite{Natsuume:2015}. Let us now define an outward normal $n^{\rho}$, pointing along the 
increasing $\rho$, as,
%%%%
\begin{equation}\label{normal1}
g_{\rho\rho}n^{\rho}n^{\rho}=1  \qquad \Rightarrow n^{\rho}=\frac{1}{\sqrt{g_{\rho\rho}}}
\end{equation} 
%%%%
where $g_{\rho\rho}$ is the metric coefficient in (\ref{met:Apos}) corresponding to the spatial coordinate $\rho$. The trace
of the extrinsic curvature, $\mathcal{K}$, can then be written as,
%%%%
\begin{equation}\label{excur1}
\mathcal{K}=n^{\rho}\frac{\partial_{\rho}\sqrt{-\gamma}}{\sqrt{-\gamma}}.
\end{equation} 
%%%%

Next, we use (\ref{dila:rd}), (\ref{normal1}) and (\ref{excur1}) to compute the difference between the third terms in the 
$\Lambda\rightarrow\infty$ limit. After a few easy steps, we note down this expression as\footnote{Notice that, (\ref{dif:3rd}) is cut-off independent. Therefore this adds a finite contribution to the free-energy. However, this boundary term (\ref{dif:3rd})
does not has a smooth $Q\rightarrow 0$ limit and is therefore valid only at finite and non-zero $Q$. Also note that, in the limit
$Q\rightarrow 0$ one can still has a finite contribution provided the temperature is also low ($\sqrt{\mu}\ll1$) such that the
ratio $\sqrt{\mu}/Q^{2}$ is small but \emph{finite}.},
%%%%
\begin{align}\label{dif:3rd}
\begin{split}
\Delta\Sigma&=\Sigma^{\text{BH}}-\Sigma^{\text{vac}} \\
&=\frac{\beta_{1}}{8}\Bigg(-14Q^{2}+\mu+\frac{\mu}{Q^{4}\mathsf{C}_{1}^{2}}
-\frac{\sqrt{\mu}\left(-4+4\mu+Q^{2}\right)}{Q^{2}\mathsf{C}_{1}(\mu-1)}
-\frac{\mu}{2\mathsf{C}_{1}^{2}Q^{4}}-\frac{\sqrt{\mu}(Q^{2}-4)}{Q^{2}\mathsf{C}_{1}}\Bigg)
\end{split}
\end{align}
%%%%

After performing a long but simple calculation, the difference in the on-shell action can be expressed as,
%%%%
\begin{align}\label{dif:act}
\begin{split}
&-\Delta S^{(os)}  \\
&=\beta_{1}\Bigg[C\Bigg\{-2\sqrt{\mu}+Q^{2}\Bigg(\frac{1}{2(\mu-1)}
+\left(1+\mathsf{C}_{1}\sqrt{\mu}\right)-\frac{\left(1+4\mu\mathsf{C}_{1}
(\mu-1)\right)}{2\sqrt{\mu}(\mu-1)}\Bigg)\Bigg\}\\
&\qquad-2Q^{2}\left(\frac{1+2\sqrt{\mu}}{1+\sqrt{\mu}}
\right)+\frac{1}{8}\Bigg(-14Q^{2}+\mu+\frac{\mu}{Q^{4}\mathsf{C}_{1}^{2}}
-\frac{\sqrt{\mu}\left(-4+4\mu+Q^{2}\right)}{Q^{2}\mathsf{C}_{1}(\mu-1)}  \\
&\qquad-\frac{\mu}{2\mathsf{C}_{1}^{2}Q^{4}}-\frac{\sqrt{\mu}(Q^{2}-4)}{Q^{2}
\mathsf{C}_{1}}\Bigg)\Bigg]
\end{split}
\end{align}
%%%%
where we have used the relation (\ref{rel:cont}).

We now analyse the behaviour of the free-energy of the configuration. In the path integral formulation, the free-energy
is given by $\displaystyle \Delta\mathcal{F}=-\beta^{-1}\log\mathcal{Z}$ where $\mathcal{Z}$ is the partition function 
and is defined as\footnote{This definition of the partition function arises from evaluating the path integral for field 
configurations close to the classical field $\phi_{cl}$ which satisfies the classical equations of motion. If $S[\phi]$ is the 
corresponding action then the path integral is dominated by fields $\phi\approx\phi_{cl}+\delta\phi$. Expanding the action 
for such fields we obtain
%%%%
\begin{equation*}
S[\phi_{cl}+\delta\phi]\simeq S[\phi_{cl}]+\frac{\delta S[\phi]}{\delta\phi}\Bigg|_{\phi_{cl}}
\delta\phi +\frac{1}{2}\frac{\delta^{2}S[\phi]}{\delta\phi^{2}}\Bigg|_{\phi_{cl}}\delta\phi^{2}
+\cdots
\end{equation*}
%%%%
Considering the variation of the fields at the boundary vanishes and neglecting the higher order terms, the 
on-shell action is approximately given by $S[\phi_{cl}]$.}$\mathcal{Z}:=e^{-\Delta S^{(os)}}$.
Thus the free-energy for the present configuration may be expressed as,
%%%%
\begin{equation}\label{ther:ener}
\Delta\mathcal{F}:=-\beta_{1}^{-1}\left(-\Delta S^{(os)}\right).
\end{equation}
%%%%

\begin{figure}[h!]
    \centering
    %\subfloat[label 1]
    {{\includegraphics[width=7cm]{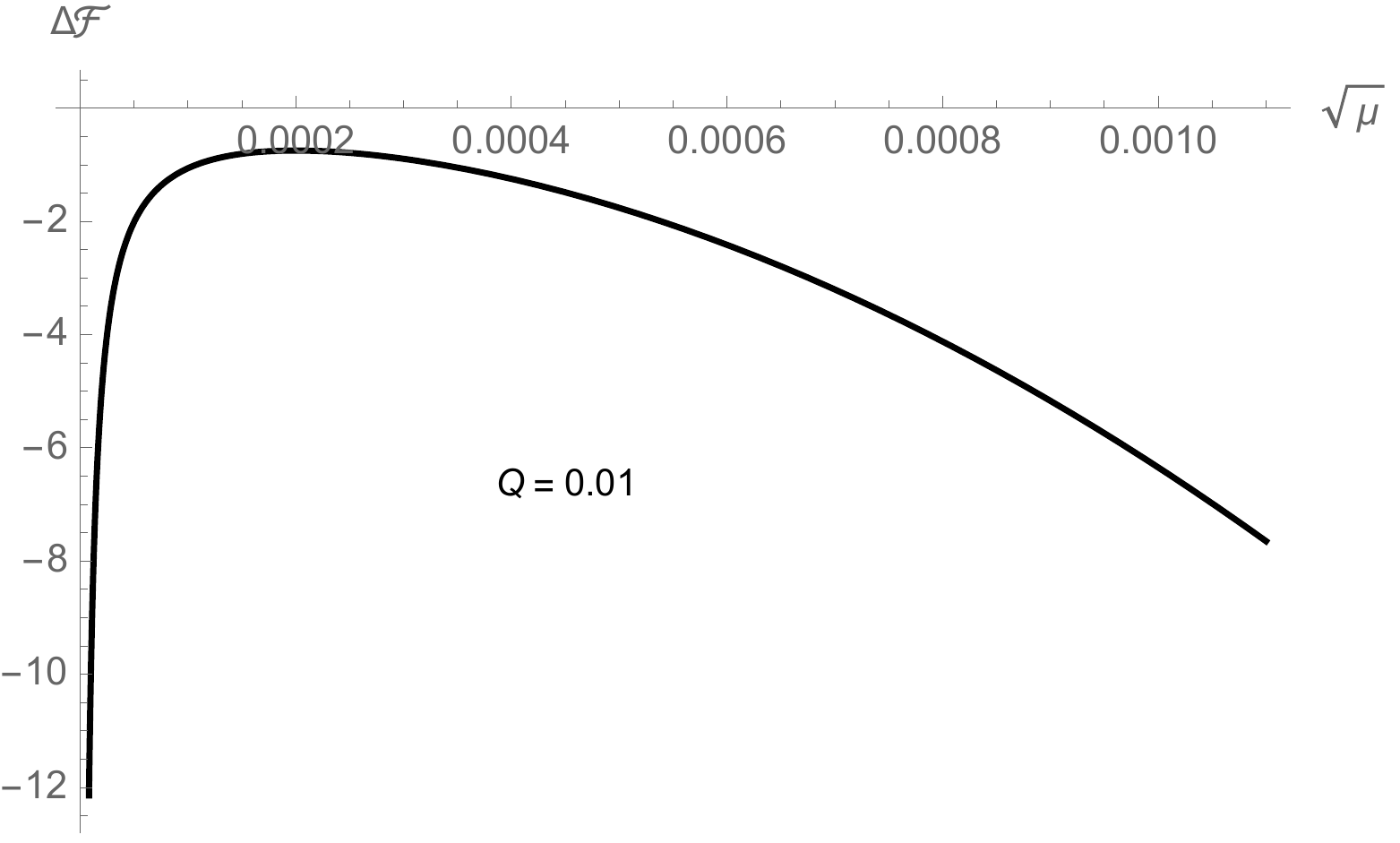} }}%
    \qquad
    %\subfloat[label 2]
    {{\includegraphics[width=7cm]{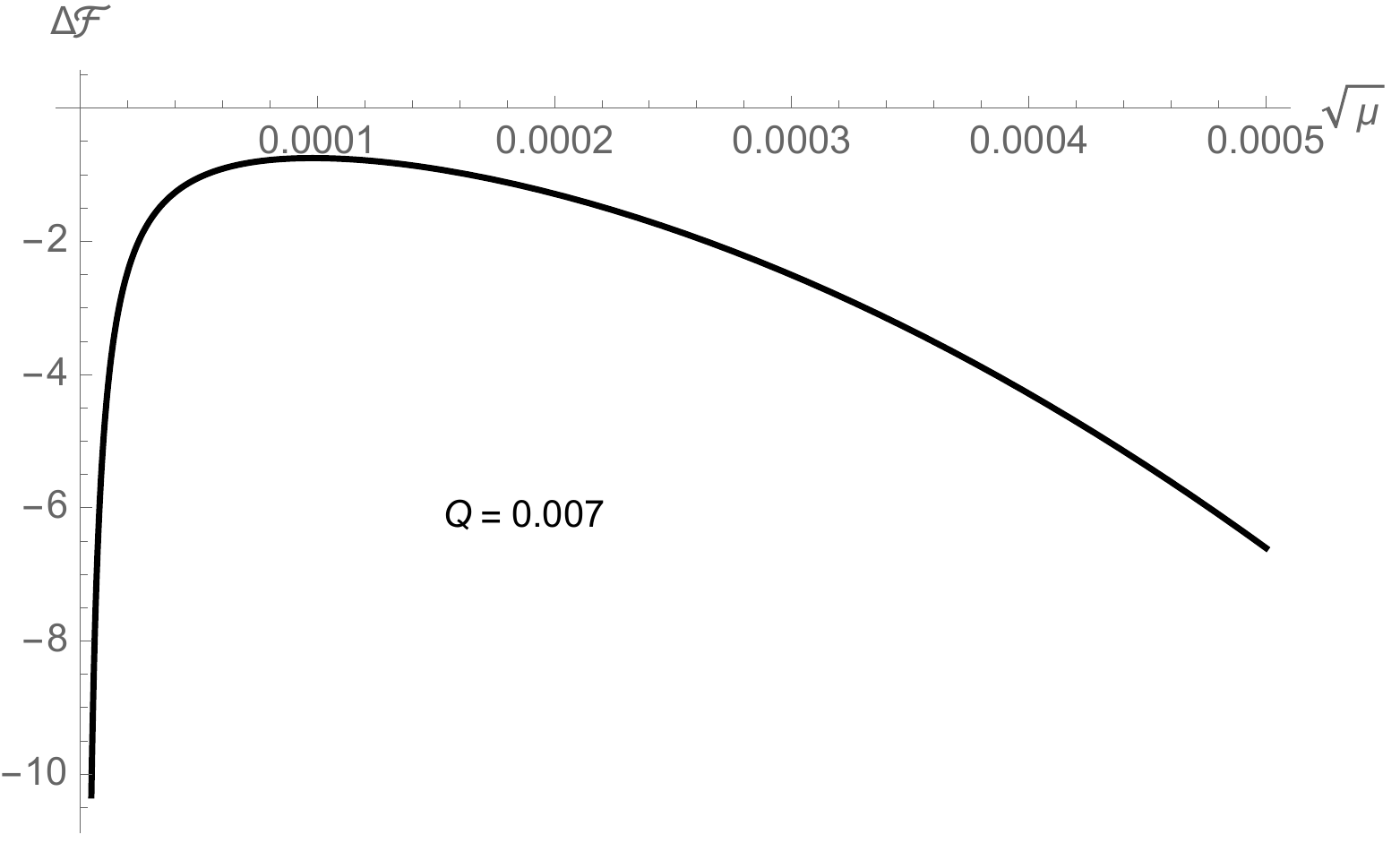} }}%
    %\qquad
    %\subfloat[label 3]
    %{{\includegraphics[width=4.5cm]{quiver_1_alpha_pp.jpg} }}
   % \caption{2 Figures side by side}%
\caption{Free energy of the black hole as a function of temperature for different values of 
charge $Q$. We have taken $\mathsf{C}_{1}=1$, $\mathsf{C}_{2}=0$ and $C=2$.}
\label{Fig:1}
\end{figure}

In Fig.\ref{Fig:1} we present the behaviour of the free-energy $\Delta\mathcal{F}$ between the black hole and the 
interpolating vacuum as a function of temperature (\ref{temp:bh}). We observe that $\Delta\mathcal{F}<0$ for all
values of the temperature $T_{H}\sim\sqrt{\mu}$. The free-energy increases asymptotically for sufficiently small
values of temperature and changes its slope at some particular temperature. Afterwards it continues to decrease.

%We observe that at low temperature and small values 
%of the charge $\Delta\mathcal{F}<0$ which implies that the black hole phase corresponding to $\mu<1$ is more stable. 
%On the other hand, as the temperature increases $\Delta\mathcal{F}$ gradually becomes positive which leads towards
%instabilities associated to $\mu>1$ branch.
%%%%
%%%%

%In our subsequent discussions we shall set $\mathsf{C}_{1}=1$, $\mathsf{C}_{2}=0$ and $C=2$. The 
%\emph{cross-over temperature}, $\sqrt{\mu}_{c}$, is obtained by setting (\ref{ther:ener}) to zero. For example,
%setting $Q=0.01$ the corresponding cross-over temperature is found to be $\sqrt{\mu}_{c}\approx 0.00352918$.

\begin{figure}[h!]
    \centering
%    \subfloat[label 1]
    {{\includegraphics[width=7cm]{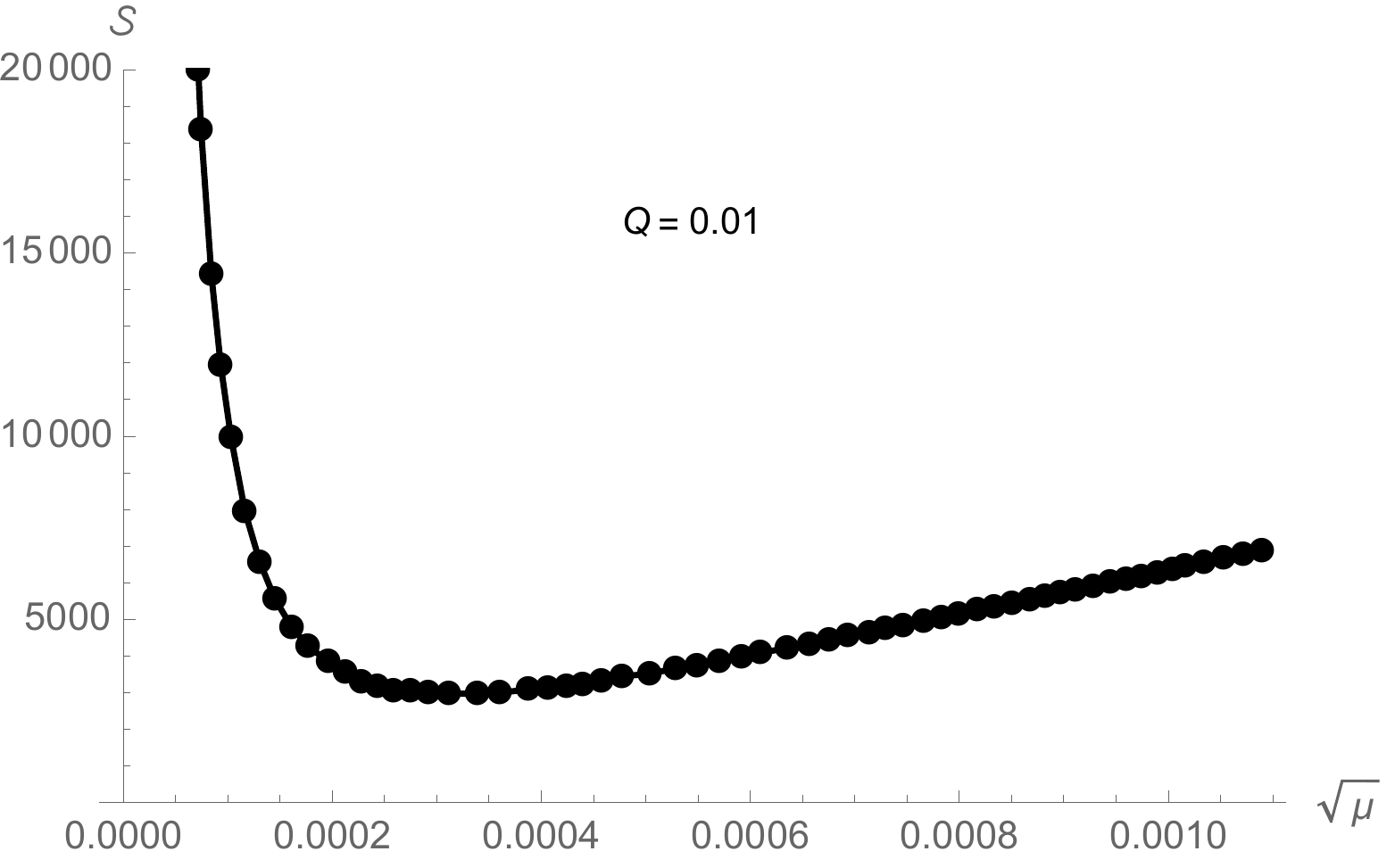} }}%
    \qquad
  %  \subfloat[label 2]
    %{{\includegraphics[width=7cm]{EntB.pdf} }}%
    %\qquad
    %\subfloat[label 3]
    {{\includegraphics[width=7cm]{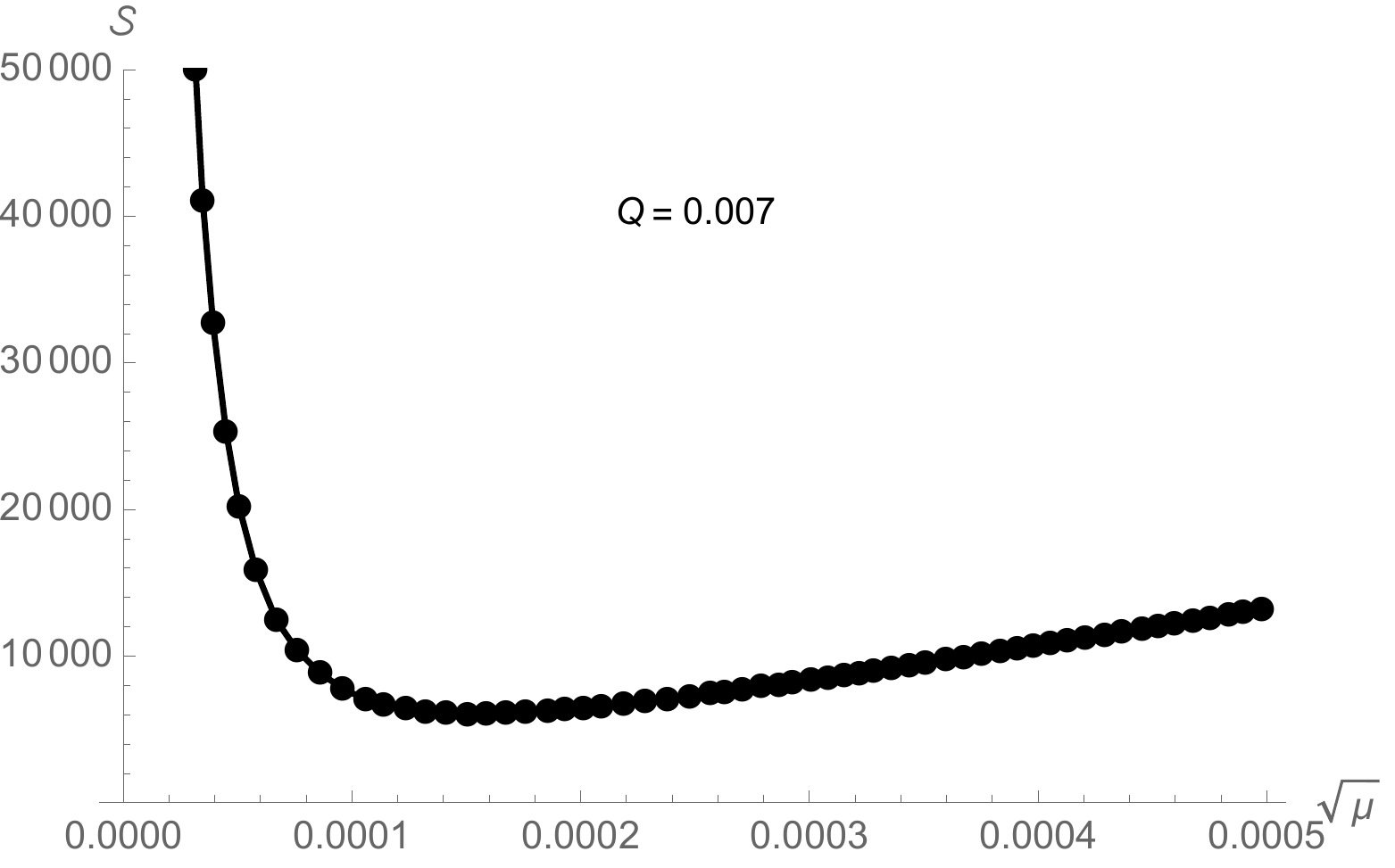} }}
   % \caption{2 Figures side by side}%
\caption{Thermodynamic entropy $\mathcal{S}=-\frac{\Delta\mathcal{F}}{\Delta\sqrt{\mu}}$ of the 
black hole as a function of temperature $\sqrt{\mu}$ for different values of charge and 
$\mathsf{C}_{1}=1$, $\mathsf{C}_{2}=0$ and $C=2$.}
\label{Fig:2}
\end{figure}

In Fig.(\ref{Fig:2}), we plot the black hole entropy $\mathcal{S}=\mathcal{S}_{W}=-\frac{\Delta\mathcal{F}}{\Delta
\sqrt{\mu}}$ against temperature $T_{H}\sim\sqrt{\mu}$. In these plots the entropy is continuous and increases smoothly 
as we lower the temperature which rules out the possibility of a first order phase transition. It is to be noted that, this 
behaviour of entropy is consistent with the Wald entropy of the black hole in (\ref{ent:bh}). This allows us to conclude that 
the phase diagram corresponding to the parameter space with $\mu<1$ is thermodynamically more preferred than that of 
the $\mu>1$ branch. We comment on the plausible implications of such phase stabilities on the dual SYK physics in the 
concluding remarks.
 
%%%%
%%%%
 
%%%%%%%%  Alignment example  %%%%%%%%

%\begin{equation}
%u(t)=\left\{\begin{array}{cc} 0 & t=0\\ 1 & t\ge0 \end{array} \right. \label{eq2}
%\end{equation} 

%%%%%%%%                                    %%%%%%%%

%%%%%%%%%%
%%%%%%%%%%
\section{Example II: Exponential coupling}\label{exp:grav}
In this section we consider the \emph{effective} 2D gravity action (\ref{action:gen}) of the following form,
%%%%
\begin{equation}\label{act:ed}
S_{ed}=\int d^{2}x\sqrt{-g}\;\left(\mathcal{R}\Phi^{2}-V(\Phi^{2})-\frac{\lambda}{4}
e^{-\Phi^{2}}F_{\mu\nu}F^{\mu\nu}\right)+\int dt \sqrt{-\gamma}\;\Phi^{2}\mathcal{K}
\end{equation}
%%%%
where $\lambda$ is the coupling constant. Also, in our subsequent analysis we shall choose the potential as 
\cite{Almheiri:2014cka},
%%%%
\begin{equation}\label{pot:ed}
V(\Phi^{2})=C-A\Phi^{2}, \qquad\qquad C,A=2.
\end{equation}
%%%%

The corresponding equations of motion are given by,
%%%%
%%%%
\begin{subequations}
\setlength{\jot}{8pt}
\begin{eqnarray}
0&=&\left(\nabla_{\mu}\nabla_{\nu}-g_{\mu\nu}\Box\right)\Phi^{2}+\frac{\lambda}{2}
e^{-\Phi^{2}}\left(F_{\mu\rho}F_{\nu}^{\;\;\rho}-\frac{1}{4}F^{2}g_{\mu\nu}\right)
-\frac{1}{2}g_{\mu\nu}V(\Phi^{2}) \label{mom:ed}\\
0&=&\mathcal{R}+\frac{\lambda}{4}e^{-\Phi^{2}}F^{2}-\frac{\partial V(\Phi)}{\partial\Phi^{2}} 
\label{som:ed}\\
0&=&\partial_{\mu}\left(\sqrt{-g}e^{-\Phi^{2}}F^{\mu\nu}\right).  \label{gom:ed}
\end{eqnarray}
\end{subequations}
%%%%

In the next step, using the definition of the metric (\ref{gauge:conf}) and the light cone coordinates $\displaystyle 
x^{\pm}\equiv (t\pm z)$ along with the ansatz (\ref{anz:gf}) we rewrite the above equations of motion as,
%%%%
%\begin{subequations}
%\setlength{\jot}{8pt}
%\begin{eqnarray}
%4\partial_{+}\partial_{-}\Phi^{2}-e^{2\omega}V(\Phi^{2})-2\lambda e^{-\Phi^{2}}
%e^{-2\omega}F_{+-}^{2}&=&0 \label{met1:ed}\\
%\partial_{\pm}\partial_{\pm}\Phi^{2}-2\left(\partial_{\pm}\omega\right)
%\left(\partial_{\pm}\Phi^{2}\right)&=&0 \label{met2:ed}\\
%4\partial_{+}\partial_{-}\omega -\frac{e^{2\omega}}{2}\frac{\partial V(\Phi^{2})}{
%\partial \Phi^{2}}-\lambda e^{-\Phi^{2}}e^{-2\omega}F_{+-}^{2}&=&0 \label{dil:ed}\\
%\partial_{\pm}\Psi&=&0 \label{gau:ed}
%\end{eqnarray}
%\end{subequations}
%%%%
%where $\displaystyle \Psi:=\sqrt{-g} e^{-\Phi^{2}}F^{+-}$. In the next step, using the definition of the light cone
%coordinates $\displaystyle x^{\pm}\equiv (t\pm z)$ and the ansatz (\ref{anz:gf}), we write equations (\ref{met1:ed})
%- (\ref{gau:ed}) in the following form
%%%%
\begin{subequations}
\setlength{\jot}{8pt}
\begin{eqnarray}
 (\Phi^{2})''+e^{2\omega}V(\Phi^{2})+\frac{\lambda}{2}e^{-\Phi^{2}}
 e^{-2\omega}A_{t}'^{2}&=&0\label{m1gc:ed}\\
(\Phi^{2})''-2\omega'(\Phi^{2})'&=&0 \label{m2gc:ed}\\
 2\omega''+e^{2\omega}\frac{\partial V(\Phi^{2})}{\partial \Phi^{2}}
+\frac{\lambda}{2}e^{-\Phi^{2}}e^{-2\omega}A_{t}'^{2}&=&0
\label{dgc:ed}\\
\Phi^{2}A_{t}''-A_{t}'\left(2\omega' +(\Phi^{2})'\right)&=&0 \label{ggc:ed}.
\end{eqnarray}
\end{subequations}
%%%%

The Maxwell field tensor is the solution to (\ref{ggc:ed}) and can be expressed as, 
%%%%
\begin{equation}\label{max:ed}
F_{zt}\equiv A_{t}'(z)=-Qe^{2\omega}e^{\Phi^{2}}.
\end{equation}
%%%%

Notice that, in the absence of charge ($Q$) the equations of motion correspond to those of the Almheiri-Polchinski
model \cite{Almheiri:2014cka}. This allows us to perform a perturbative expansion of the dilaton as well as the metric
as done before and find the corresponding vacuum as well as the (charged) black hole solutions.

\subsection{Interpolating vacuum solution}\label{vsol:ed}
In the following, we note down the $\mathcal{O}(Q^{2})$ equations of motion corresponding to the metric as well as
the dilaton,
%%%%
\begin{align}
0&=2\omega_{(1)}''-4e^{2\omega_{(0)}}\omega_{(1)}+\frac{\lambda}{2}e^{2\omega_{(0)}}
e^{\Phi_{(0)}^{2}} \label{omv:ed}\\
0&=\left(\Phi_{(1)}^{2}\right)''-2e^{2\omega_{(0)}}\Phi_{(1)}^{2}+4e^{2\omega_{(0)}}
\left[\omega_{(1)}\left(1-\Phi_{(0)}^{2}\right)+\frac{\lambda}{8}e^{\Phi_{(0)}^{2}}\right] 
\label{phiv:ed}
\end{align}
%%%%
while $\mathcal{O}(1)$ solutions are given by (\ref{vsol:w0}) and (\ref{vsol:p0}).

The solution to (\ref{omv:ed}) can be found as,
%%%%
\begin{equation}\label{om1v:ed}
\omega_{(1)}(z)=\frac{\mathsf{C}_{5}}{z}-\frac{\lambda}{24z}\left(-2ze^{1+1/2z}
\left(1+2z-4z^{2}\right)+e\;\text{Ei}\left(\frac{1}{2z}\right)\right)
\end{equation}
%%%%
where the exponential integral function is given by,
%%%%
\begin{equation}\label{int:exp}
\text{Ei}(z)=-\mathcal{P}\int_{-z}^{\infty}dt \frac{e^{-t}}{t}.
\end{equation}
%%%%

Finally, we express the metric (\ref{gauge:conf}) as,
%%%%
\begin{align}\label{metv:ed}
%\begin{split}
ds^{2}=e^{2\omega_{(0)}(z)}\left(1+2Q^{2}\omega_{(1)}(z)\right)\left(-dt^{2}+dz^{2}\right)
%\end{split}
\end{align}
%%%%

The behaviour of the metric (\ref{gauge:conf}) in the IR ($z\rightarrow \infty$) is obtained as,
%%%%
\begin{equation}\label{IR:ed}
e^{2\omega(z)}\Big|_{z\rightarrow \infty}\simeq -\frac{2}{3}e\lambda Q^{2}
+\frac{1}{z^{2}}\left(1+\frac{1}{4}e\lambda Q^{2}\right)+\mathcal{O}\left(
\frac{1}{z^{3}}\right)
\end{equation}
%%%%
which clearly indicates that the IR behaviour of the geometry is AdS$_2$. On the other hand, near the boundary
($z\approx 0$) the metric (\ref{gauge:conf}) behaves as,
%%%%
\begin{equation}\label{UV:ed}
e^{2\omega(z)}\Big|_{z\approx 0}\simeq \frac{2Q^{2}\mathsf{C}_{5}}{z^{3}}
+\frac{1}{z^{2}}+e^{1/2z}\frac{1}{\mathcal{O}(z)}+\mathcal{O}(1)
\end{equation}
%%%%
whose leading order term is $\sim 1/z^{3}$ which is a signature of an asymptotically Lifshitz space-time with
dynamical scaling exponent $z_{\text{dyn}}=3/2$. Thus the geometry interpolates between AdS$_2$ in the IR and 
Lifshitz$_2$ in the UV. This observation is similar in spirit to what we have found in the earlier example.

\subsection{Black hole solution}\label{lifsol:ed}
In order to obtain the charged black hole solution we recall the corresponding zeroth order solutions (\ref{solper:phi0}) 
and (\ref{solper:om0}) and substitute them into (\ref{phiv:ed}) which finally yields the first order correction to the metric,
%%%%
\begin{align}\label{solw1:ed}
\begin{split}
&\omega_{(1)}(\rho)\\
&=\mathsf{C}_{6}\frac{\rho}{\sqrt{\mu}}+\mathsf{C}_{7}\Bigg[
-1+\frac{\rho}{\sqrt{\mu}}\cdot\log\sqrt{\frac{\sqrt{\mu}+\rho}{\sqrt{\mu}-\rho}}\;
\Bigg] \\
&-\frac{\lambda e^{1-\sqrt{\mu}}}{8\mu\sqrt{\mu}}\Big[-2\sqrt{\mu}
e^{\rho+\sqrt{\mu}}+\rho e^{2\sqrt{\mu}}\left(\sqrt{\mu}-1\right)\text{Ei}(\rho-
\sqrt{\mu})+\rho\left(\sqrt{\mu}+1\right)\text{Ei}(\rho+\sqrt{\mu})\Big]
\end{split}
\end{align}
%%%%
where $\mathsf{C}_{6}$ and $\mathsf{C}_{7}$ are arbitrary integration constants and $\text{Ei}(z)$ is given
in (\ref{int:exp}). Notice that, in finding the above solution we have used the change in coordinate (\ref{coor:new}).

Finally, using (\ref{solper:om}) we note down the metric (\ref{gauge:conf}),
%%%%
\begin{align}\label{metbh:ed}
\begin{split}
ds^{2}&=e^{2\omega}\left(-dt^{2}+dz^{2}\right)\\
%&=\frac{4\mu}{\sinh^{2}2\sqrt{\mu}z}\left(1+2Q^{2}\omega_{(1)}^{BH}(z)\right)
%\left(-dt^{2}+dz^{2}\right)\\
&=4\left(\rho^{2}-\mu\right)\left(1+2Q^{2}\omega_{(1)}^{BH}(\rho)\right)\left(-dt^{2}+
\frac{d\rho^{2}}{4\left(\rho^{2}-\mu\right)^{2}}\right).
\end{split}
\end{align}
%%%%

Clearly, in this coordinates the horizon of the black hole is located at $\rho=\sqrt{\mu}$. On the other hand, the 
behaviour of the metric (\ref{metbh:ed}) near the boundary $\rho\rightarrow\infty$ may be found as,
%%%%
\begin{equation}\label{asym:ed}
e^{2\omega}(\rho)\Big|_{\rho\rightarrow\frac{1}{\Delta}}\simeq \frac{e^{\frac{1}{\Delta}}}{
\mathcal{O}(\Delta)}+\frac{8Q^{2}\mathsf{C}_{6}}{\sqrt{\mu}\Delta^{3}}+\frac{4}{\Delta^{2}}
+\frac{1}{\mathcal{O}(\Delta)}
\end{equation}
%%%%
where we have set $\mathsf{C}_{7}=0$ in order to obtain \emph{physical} boundary conditions in the asymptotic 
limit of the metric. Referring to (\ref{asym:ed}), we may conclude that the geometry is asymptotically Lifshitz near the 
boundary of the space-time with dynamical exponent $z_{\text{dyn}}=3/2$. This observation is similar to that observed 
in Section \ref{sol:lif2}.

\subsection{Free-energy and phase stability}\label{phase:ed}
In order to understand the underlying phase structure associated with the black hole solution (\ref{metbh:ed}) we 
follow the same line of analysis as in Section \ref{phtran}. The on-shell action for the present model may be calculated 
as,
%%%%
\begin{align}\label{actos:ed}
\begin{split}
-S^{(os)}_{ed}&=C\int d^{2}x\sqrt{-g}+\frac{\lambda}{2}\int d^{2}x\;
\partial_{\mu}\left(\sqrt{-g}e^{-\Phi^{2}}\left(1+\Phi^{2}\right)
F^{\mu\nu}A_{\nu}\right) \\
&\qquad -\frac{\lambda}{2}\int d^{2}x\;\partial_{\mu}
\left(\sqrt{-g}e^{-\Phi^{2}}\Phi^{2}F^{\mu\nu}\right)A_{\nu}
-\int dt \;\sqrt{-\gamma}\mathcal{K}\Phi^{2}.
\end{split}
\end{align}
%%%%

The Abelian $1$-form in (\ref{actos:ed}) may be replaced as,
%%%%
\begin{equation}\label{gauge:ed}
A_{t}=-Q\int dz\; e^{2\omega_{(0)}}e^{\Phi_{(0)}^{2}}
\left(1+\mathcal{O}(Q^{2})\right)
\end{equation} 
%%%%
where the exponentials in the integrand stand for the zeroth order solutions to the metric and the dilaton. 

If we now substitute (\ref{max:ed}) and (\ref{gauge:ed}) into (\ref{actos:ed}) it is easy to check that the second and the
third terms in the on-shell action (\ref{actos:ed}) exactly cancel each other. On the other hand, the difference in the volume 
terms may be schematically expressed as,
%%%%
\begin{align}\label{dvol:ed}
\begin{split}
\Delta\left(\text{Vol}\right)&=C\left(\left(\text{Vol}\right)^{\text{BH}}
-\left(\text{Vol}\right)_{\text{Int}}^{\text{vac}} \right)   \\
&=-C\beta_{1}\Big[\mathcal{D}_{1}-\mathcal{D}_{2}^{(\sqrt{\mu})}-
\mathcal{D}_{3}\frac{\beta_{0}}{\beta_{1}}\Big]
\end{split}
\end{align}
%%%%
whose finite contribution may be expressed as,
%%%%
\begin{equation}\label{vfin:ed}
\mathcal{D}_{2}^{(\sqrt{\mu})}=\frac{1}{2\sqrt{\mu}}\Big[e^{1+\sqrt{\mu}}
Q^{2}\lambda+4\mu\left(1+\mathsf{C}_{6}Q^{2}\right)\Big].
\end{equation}
%%%%

Finally, using the definition (\ref{ther:ener}) the corresponding free-energy can be expressed as\footnote{For the present 
model, the Gibbons-Hawking-York boundary term in (\ref{actos:ed}) does not provide any finite contribution to the 
free-energy.}
%%%%
\begin{equation}\label{free:ed}
\Delta\mathcal{F}=-\frac{1}{\pi}\Big[e^{1+\sqrt{\mu}}Q^{2}
\lambda+4\mu\left(1+\mathsf{C}_{6}Q^{2}\right)\Big]
\end{equation}
%%%%
where $\beta_{1}\equiv 1/T_{H}=\pi/\sqrt{\mu}$ is the inverse Hawking temperature of the black hole, which is easily 
calculated by substituting the metric components of (\ref{metbh:ed}) in (\ref{temp:bh}).
%%%%
\begin{figure}[h!]
    \centering
    %\subfloat[label 1]
    {{\includegraphics[width=10cm]{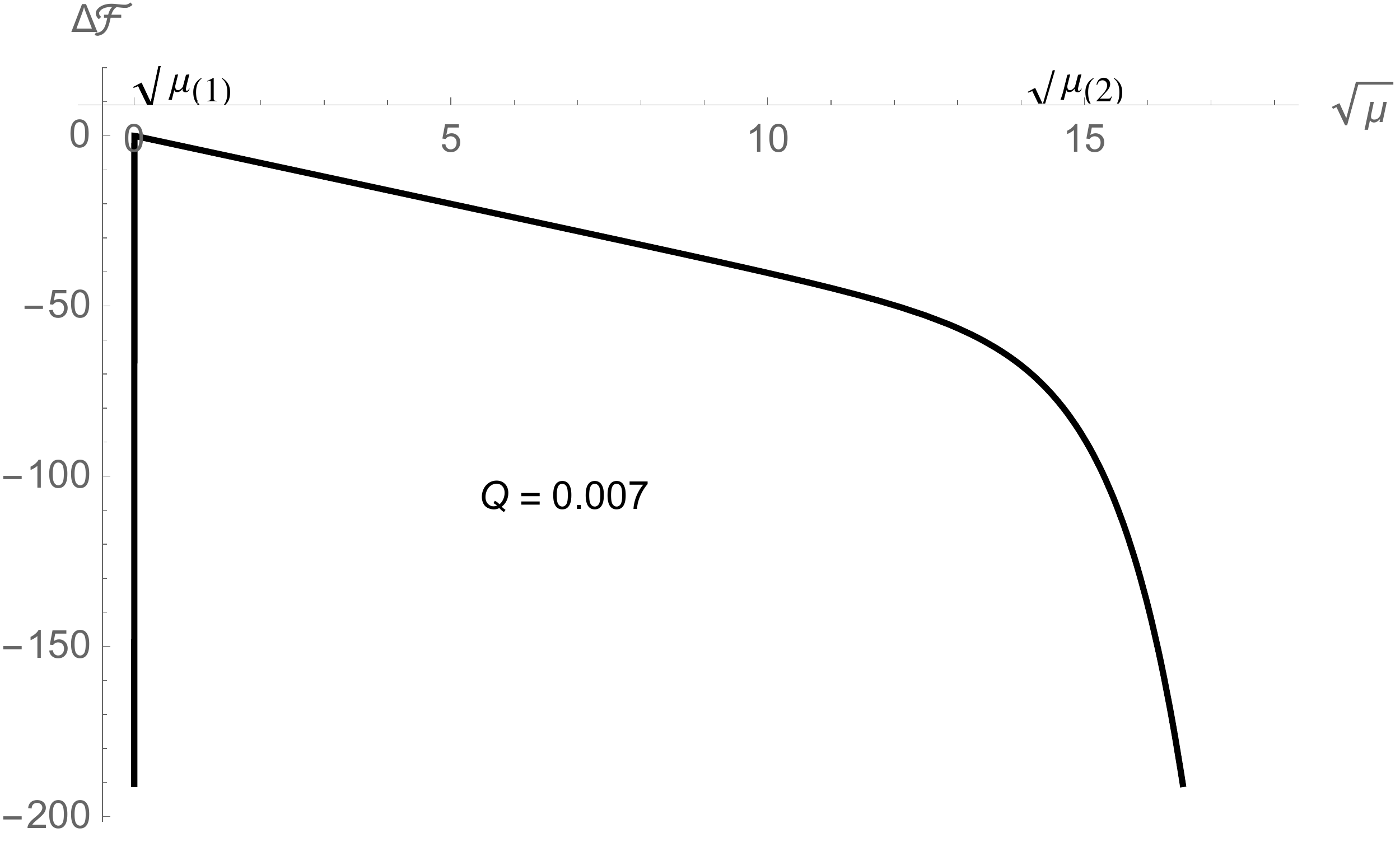} }}%
   % \qquad
    %\subfloat[label 2]
    %{{\includegraphics[width=4.5cm]{quiver_1_alpha_p.jpg} }}%
    %\qquad
    %\subfloat[label 3]
    %{{\includegraphics[width=4.5cm]{quiver_1_alpha_pp.jpg} }}
   % \caption{2 Figures side by side}%
\caption{Free energy of the black hole as a function of temperature for $Q=0.007$. 
We have taken $\mathsf{C}_{6}=1$ and $\mathsf{C}_{7}=0$.}
\label{Fig:3}
\end{figure}
%%%%

%%%%
\begin{figure}[h!]
    \centering
%    \subfloat[label 1]
    {{\includegraphics[width=7cm]{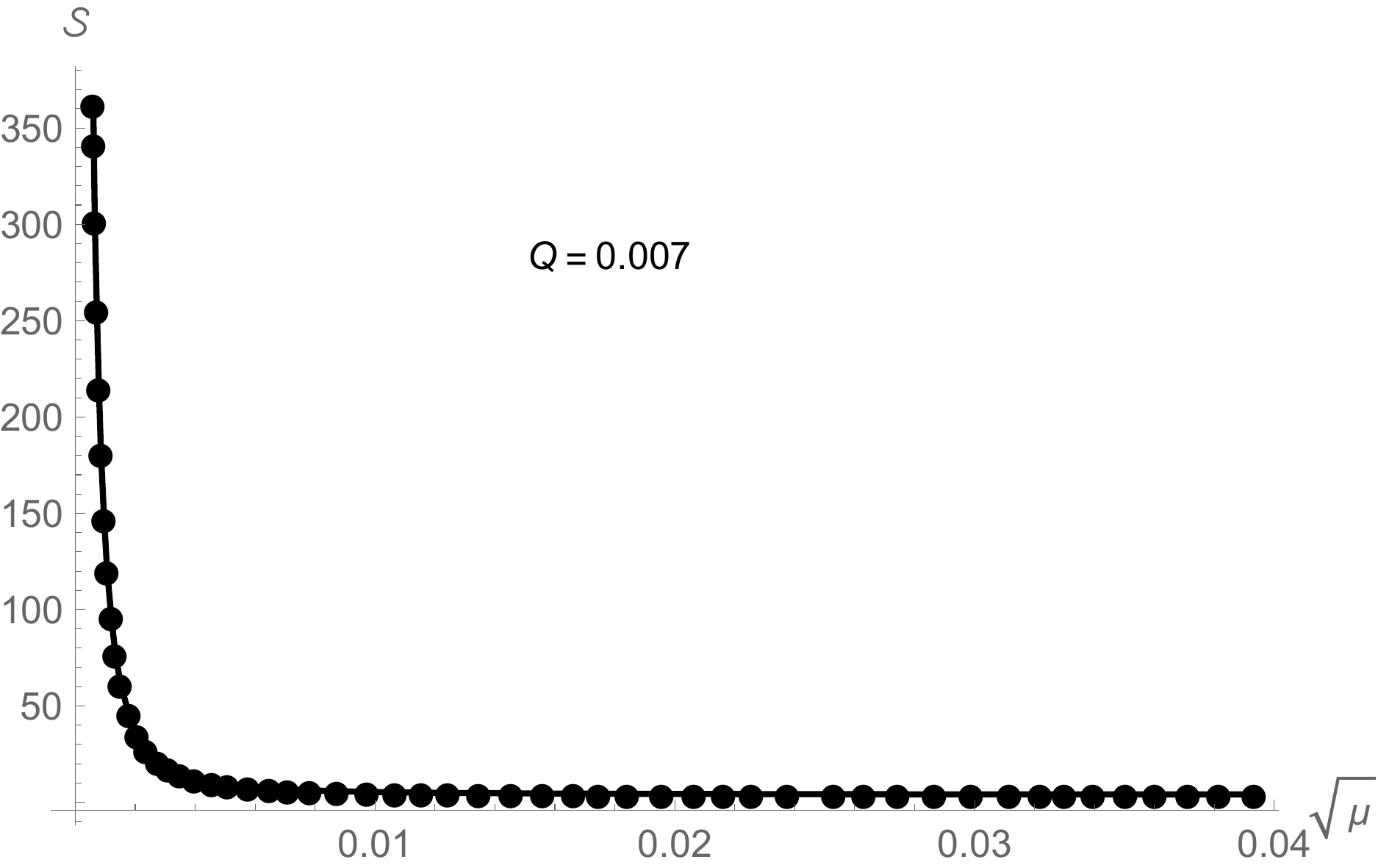} }}%
    \qquad
  %  \subfloat[label 2]
    {{\includegraphics[width=7cm]{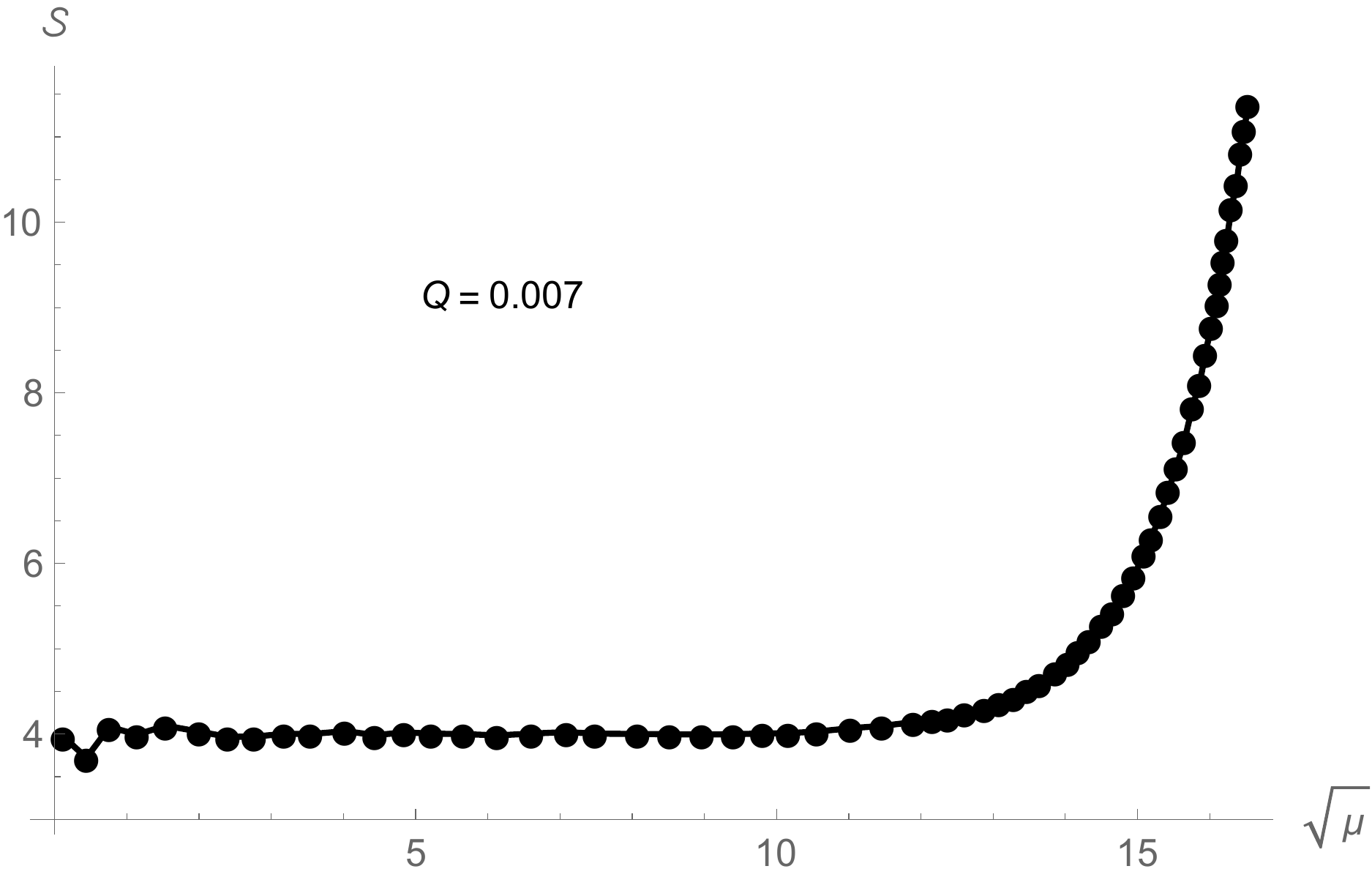} }}%
 %   \qquad
    %\subfloat[label 3]
  %  {{\includegraphics[width=7cm]{entplot4.pdf} }}
   % \caption{2 Figures side by side}%
\caption{Thermodynamic entropy $\mathcal{S}=-\frac{\Delta\mathcal{F}}{\Delta\sqrt{\mu}}$ of the 
black hole as a function of temperature $\sqrt{\mu}$. The left panel corresponds to the left-half 
($\sqrt{\mu}\in(0,0.04)$) of the corresponding curve in Fig.\ref{Fig:3} while the right panel shows 
the remaining right-half of the same curve. Here $Q=0.007$ and $\mathsf{C}_{6}=1$.}
\label{Fig:4}
\end{figure}
%%%%
We now plot the free-energy ($\Delta\mathcal{F}$) of the black hole against the temperature, $T_{H}\sim\sqrt{\mu}$
(see Fig.\ref{Fig:3}). Further in Fig.\ref{Fig:4}, we show the changes in entropy $\mathcal{S}=S_{W}$ ($=- 
\Delta\mathcal{F}/\Delta\sqrt{\mu}$) against the temperature $T_{H}\sim\sqrt{\mu}$. From this plot it is evident that 
there is no first order phase transition as the $\mathcal{S}$-$\sqrt{\mu}$ plot is continuous. We further observe that, this
behaviour of entropy is consistent with that of the Wald entropy of the corresponding black hole which can be easily
computed using (\ref{solper:phi}) and (\ref{dila:rd}) and is given by,
%%%%
\begin{equation}\label{wald:ed}
S_{W}=4\pi\left[1+\sqrt{\mu}+Q^{2}\frac{\lambda e^{1+\sqrt{\mu}}
+4\mathsf{C}_{6}\mu}{4\sqrt{\mu}}\right].
\end{equation}
%%%%

From Fig.\ref{Fig:3} we notice two \emph{turning points}, $\sqrt{\mu}_{(1)}$ and $\sqrt{\mu}_{(2)}$ ($\sqrt{\mu}_{(1)} 
<\sqrt{\mu}_{(2)}$) at which the sign of the slope changes. This behaviour is reflected in the corresponding entropy 
plots in Fig.\ref{Fig:4} in which the entropy increases asymptotically both for $\sqrt{\mu}<\sqrt{\mu}_{(1)}$ and 
$\sqrt{\mu}>\sqrt{\mu}_{(2)}$. In the window $\Delta\sqrt{\mu}\sim\sqrt{\mu}_{(2)}-\sqrt{\mu}_{(1)}$ the entropy does 
not change considerably. In this window the system falls in the minimum entropy state whose plausible interpretation 
in terms of SYK degrees of freedom are discussed in the concluding remarks.

%%%%%%%%%%%%%%%%%%%%%%%%
%%%%%%%%%%%%%%%%%%%%%%%%

\section{Concluding remarks}\label{conclusions}
In the present paper, we have proposed models of charged solutions within the framework of $1+1$ D Jackiw-Teitelboim 
(JT) gravity. Based on our model computations, we have shown that in the presence of non-trivial couplings between the 
$U(1)$ gauge field and the dilaton the asymptotic geometries get substantially modified  both for the vacuum as well as 
the black hole solutions. In both the examples, the vacuum solutions interpolate between Lifshitz$_2$ in the UV to 
AdS$_2$ in the IR. On the other hand, the black hole solutions turn out to be asymptotically Lifshitz$_2$ with 
$z_{\text{dyn}}=3/2$.

We have further analysed the stability of black holes in both the models and observed a \textit{universal} feature in the free-energy
and therefore the entropy of the system. In the first model we have considered the quadratic coupling and in the second model the 
exponential coupling of the dilaton to the gauge field. In both the cases, at sufficiently low temperatures, we observed a turning point after which the free-energy falls-off sharply (Fig.\ref{Fig:1} and Fig.\ref{Fig:3}) which corresponds to an increase in the entropy
below this temperature (Fig.\ref{Fig:2} and Fig.\ref{Fig:4}).

The existence of minimal entropy at low temperatures could be interpreted as the formation of the Bose-Einstein like 
condensate\footnote{Notice that the classical solutions in the JT gravity correspond to large $N$ dynamics in the dual SYK
model. Therefore, one should interpret the charged condensate as a \emph{classical} (large $N$) analogue 
of the BEC-like phenomena \cite{Staliunas:2000a}-\cite{Utyuzh:2007ct} in the dual SYK picture where quantum fluctuations are suppressed because of $1/N$ corrections.}(BEC) in the dual SYK model which possibly leads towards superfluid instabilities at low 
temperatures and finite density \cite{Tiley:ss}. We hope to clarify some of these issues from the perspective of the dual SYK physics in the near future.

\section*{Acknowledgments}
The work of A.L is supported by the Chilean FONDECYT project No. 3190021. D.R is indebted to the authorities 
of IIT Roorkee for their unconditional support towards researches in basic sciences. Both the authors would like 
to thank Hemant Rathi, Jitendra Pal and Dr. Arup Samanta for useful discussions. Special thanks to Jakob Salzer 
for his valuable comments on the manuscript.  
%%%%%%%%%%%%%%%%%%%%%%%%%%%%%%%%%%%%%%%%%%%%%%%%%%%%%%%%%%%%%%%%%%%%%%%%%%%%%%%%%%%%%%%%%%%%%%%%
\appendix
%%%%%%%%%%%%%%%%%%%%%%%%%%%%%
%%%%%%%%%%%%%%%%%%%%%%%%%%%%%
\section{A note on dimensional reduction}\label{dim:redn}
In this section, we propose a dimensional reduction procedure in order to show that our $1+1$ D dilaton gravity models 
(\ref{act:charged}) and (\ref{act:ed}) are indeed effective models of a higher dimensional gravity theory.

Let us consider the following $2+1$ D Einstein-dilaton gravity,
%%%%
\begin{equation}\label{act:3D}
S_{3D}=\int d^{3}x\sqrt{-g^{(3)}}\left(\mathcal{R}^{(3)}+\chi(\Psi)+
\gamma\left(\partial_{z}\Psi\right)\left(\partial^{z}\Psi\right)\right)
\end{equation}
%%%%
where $\mathcal{R}^{(3)}$ is the three dimensional Ricci scalar and $\chi(\Psi)$ is the dilaton potential which includes
the cosmological constant as we see below. Here $\Psi\equiv\Phi^{2}$ of the original analysis.

In order to obtain a $1+1$ D effective action, we dimensionally reduce (\ref{act:3D}) along the compact direction $\theta$,
%%%%
\begin{equation}\label{ansz:kk}
ds_{(3)}^{2}=ds_{(2)}^{2}+e^{-2\beta\Psi}\left(d\theta+\tilde{A}_{a}dx^{a}\right)^{2}
\end{equation}
%%%%
where $ds_{(2)}^{2}$ is the usual AdS$_2$ metric (\ref{gauge:conf}) with $e^{2\omega}=1/z^{2}$. 

Notice that, in (\ref{ansz:kk}) the indices $a,b$ run over the uncompactified directions. Also, we have identified $\theta
\sim\theta+2\pi$ and assumed an $U(1)$ symmetry for the dilaton field $\Psi$. The gauge fields $\tilde{A}_{a}$ in 
(\ref{ansz:kk}) are known as the Kaluza-Klein vectors. In the subsequent analysis, we choose to work with the ansatz 
(\ref{anz:gf}).  

In the next step, we wish to calculate the $2+1$ D Ricci scalar $\mathcal{R}^{(3)}$ which is related to the $1+1$ D 
Ricci scalar $\mathcal{R}^{(2)}$ as,
%%%%
\begin{equation}\label{rel:ricci}
\mathcal{R}^{(3)}=\mathcal{R}^{(2)}-\frac{1}{4}e^{-2\beta\Psi}\tilde{F}_{\mu\nu}^{2}
+2\beta\left(\partial_{z}\partial^{z}\Psi\right)-2\beta^{2}\left(\partial_{z}\Psi\right)
\left(\partial^{z}\Psi\right).
\end{equation}
%%%%

We now find a relation between the determinants of the two metrics as,
%%%%
\begin{equation}\label{rel:det}
\sqrt{-g^{(3)}}=\sqrt{-g^{(2)}}e^{-\beta\Psi}.
\end{equation}
%%%%

If we now substitute (\ref{rel:ricci}) and (\ref{rel:det}) in (\ref{act:3D}), the $2+1$ D action reduces to the following
$1+1$ D form,
%%%%
\begin{align}\label{act:red}
\begin{split}
S_{2D}&=\int d^{2}x\sqrt{-g^{(2)}}e^{-\beta\Psi}\left[\mathcal{R}^{(2)}
+\chi(\Psi)-\frac{1}{4}e^{-2\beta\Psi}\tilde{F}_{\mu\nu}^{2}\right]   
%&\simeq\mathcal{V}_{\theta}\int d^{2}x\sqrt{-g^{(2)}}\left[\beta\Phi\mathcal{R}^{(2)}+\Lambda
%(1+\beta\Phi)-\frac{1}{4}e^{3\beta\Phi}\tilde{F}_{\mu\nu}^{2}\right]
\end{split}
\end{align}
%%%%
where we set $\beta=\pm\sqrt{\gamma/2}$ and $\gamma=2$. 

If we redefine $\Psi\rightarrow \log\Psi$ and use $\beta=-1$ in (\ref{act:red}) we recover an action which is similar in
spirit to the action (\ref{act:charged}) corresponding to the Model I. Notice that, in order to obtain the desired form of
the potential one must set $\chi=\left(A-C/\Psi\right)$ where $A$ plays the role of cosmological constant in the original
$2+1$ D gravity model (\ref{act:3D}). On the other hand, in the limit $\Psi\ll 1$ we obtain an action
similar to (\ref{act:ed}) which corresponds to that of Model II. In this case we set $\beta=-1$ and $\chi\approx 
-C+(C+A)\Psi$.

%%%%%%%%%%%%%%%%%%%%%%%%%%%%%
%%%%%%%%%%%%%%%%%%%%%%%%%%%%%
\section{ Black hole solution with $A<0$} \label{sols:neg}

In this appendix, we discuss the metric solution corresponding to our first model (\ref{act:charged}) while considering 
the linear dilaton potential as $U(\Phi^{2})=C+A\Phi^{2}$. In order to simplify the calculations we choose $A,C=2$.

Using the perturbation expansion (\ref{solper:phi}) and (\ref{solper:om}), we write down the equation of motion
(\ref{c:steom}) upto leading order in the expansion as,
%%%%
\begin{align}
\mathcal{O}(Q^{0}):\qquad 0&=2\omega_{(0)}''(z)+2e^{2\omega_{(0)}(z)}\label{per:zero} \\
\mathcal{O}(Q^{2}):\qquad 0&=2\omega_{(1)}''(z)+4e^{2\omega_{(0)}(z)}\omega_{(1)}(z)-
\frac{e^{2\omega_{(0)}(z)}}{(\Phi_{(0)}^{2})^{3}}.  \label{per:first}
\end{align}
%%%%

Similarly, substituting (\ref{solper:phi}) and (\ref{solper:om}) in the dilaton equation of motion (\ref{a:steom}) we 
obtain,
%%%%
\begin{align}
\mathcal{O}(Q^{0}):\qquad 0&=\left(\Phi_{(0)}^{2}\right)''+2e^{2\omega_{(0)}}\left(1+
\Phi_{(0)}^{2}\right) \label{Q0min:phi}\\
\mathcal{O}(Q^{2}):\qquad 0&=\left(\Phi_{(1)}^{2}\right)''+2e^{2\omega_{(0)}}
\Phi_{(1)}^{2}+4e^{2\omega_{(0)}}\left[\omega_{(1)}\left(1+\Phi_{(0)}^{2}\right)+
\frac{1}{8}\left(\Phi_{(0)}^{2}\right)^{-2}\right]. \label{Q1min:phi}
\end{align}
%%%%

The zeroth order equations (\ref{per:zero}) and (\ref{Q0min:phi}) have the following solutions, 
%\footnote{Note that, $e^{2 \omega_{(0)}}=\frac{4}{\cosh^{2} 2z}$ is also a solution to 
%(\ref{per:zero}). This arises from the scale invariance of the theory, namely we can scale $z$ 
%by any $\lambda\in\mathbb{R}$ and the theory still remains the same.}
%%%%
\begin{align}
e^{2 \omega_{(0)}}&=\frac{4\mu}{\cosh^{2} 2\sqrt{\mu}z}  \label{solmin:w0} \\
\left(\Phi_{(0)}^{2}\right)&=-\left(1+\sqrt{\mu}\tanh 2\sqrt{\mu}z\right)  \label{solmin:p0}
\end{align}
%%%%

Notice that, in obtaining the solution (\ref{solmin:p0}) we have set $\bar{C}=-1$ and $\tilde{C}=-\frac{1}{2}$ 
in (\ref{dila:rd}).
%\footnote{It is to be noted that if we keep the constant $\tilde{C}$ in the solution 
%the asymptotic value of the dilaton does not change. It always appears as $\tilde{C}^{2}$ 
%and hence the sign is not significant.}  

Now, using (\ref{solper:phi}) and (\ref{solper:om}) in (\ref{per:first}), the leading order solution $\omega_{(1)}^{BH}$ 
can be expressed as,
%%%%
\begin{align}\label{sol:om1}
\begin{split}
&\omega_{(1)}^{BH}(\rho)  \\
&=\frac{\rho\mathsf{C}_{8}}{\sqrt{\mu}}+\mathsf{C}_{9}\left[-1+\frac{\rho}{\sqrt{\mu}}
\tanh^{-1}\left(\frac{\rho}{\sqrt{\mu}}\right)\right]+\frac{1}{8{\mu-1}^{2}\mu^{3/2}(1+\rho)} \\
&\quad\times\Big\{-2\sqrt{\mu}(\mu-1)(-1+\mu-\rho)+\rho(1+\rho)\Big[4\mu^{3/2}\cdot\log(1+\rho)\\
&\qquad+\left(-1+3\mu-2\mu^{3/2}\right)\log(\rho-\sqrt{\mu})+\left(1-3\mu-2\mu^{3/2}\right)
\log(\rho+\sqrt{\mu})\Big]\Big\}.
\end{split}
\end{align}
%%%%
where $\mathsf{C}_{8}$, $\mathsf{C}_{9}$ are constants of integration.In writing (\ref{sol:om1}) we have made the 
following change in the spatial coordinate:
%%%%
\begin{equation}
\label{trans:coord}
z\longrightarrow\frac{1}{2\sqrt{\mu}}\tanh^{-1}\left(\frac{\rho}{\sqrt{\mu}}\right).
\end{equation}
%%%%

Finally, using (\ref{solper:om}) and (\ref{solmin:w0}), the metric (\ref{gauge:conf}) corresponding to the black hole
can be written as, 
%%%%
\begin{align}
\label{trans:metric}
\begin{split}
ds^{2}=4(\mu-\rho^{2})\left(1+2Q^{2}\omega_{(1)}(\rho)\right)
\left(-dt^{2}+\frac{d\rho^{2}}{4(\mu-\rho^{2})^{2}}\right).
\end{split}
\end{align} 
%%%%

Clearly, the horizon of the black hole (\ref{trans:metric}) is located at $\rho=\sqrt{\mu}$. However, from the structure
of the solution (\ref{solmin:w0}) the position of the boundary of the space-time (\ref{trans:metric}) is not quite apparent.

The Hawking temperature corresponding to the above black hole (\ref{trans:metric}) can be written down using the 
formula (\ref{temp:bh}) and is given by,
%%%%
\begin{equation}\label{ht:apos}
T_{H}=\frac{\sqrt{\mu}}{2\pi}.
\end{equation}
%%%%

The corresponding Wald entropy can be expressed as,
%%%%
\begin{equation}\label{ent:wald}
S_{W}=4\pi\left[1+\sqrt{\mu}+Q^{2}\frac{1+4\mathsf{C}_{8}\mu|\mu-1|}{4\sqrt{\mu}|\mu-1|}\right].
\end{equation}
%%%%
Notice that, in writing (\ref{ent:wald}) we have used (\ref{solper:phi}) and (\ref{dila:rd}).

%%%%%%%%%%%%%%%%%%%%%%%%%%
%%%%%%%%%%%%%%%%%%%%%%%%%%
%%%%%%%%%%%%%%%%%%%%%%%%%%%%%%%%%%%%%%%%%%%%%%%%%%%%%%%%%%%%%%%%%%%%%%%%%%%%%%%%%%%

\end{document}